# Limits of effective material properties in the context of an electromagnetic tissue model*

Kevin Jerbic[1,2], Kevin Neumann[1], Jan Taro Svejda[1], Benedikt Sievert[1], Andreas Rennings[1] and Daniel Erni[1]

*Abstract*— **Most calibration schemes for reflection-based tissue spectroscopy in the mm-wave/THz-frequency range are based on homogenized, frequency-dependent tissue models where macroscopic material parameters have either been determined by measurement or calculated using effective material theory. However, as the resolution of measurement at these frequencies captures the underlying microstructure of the tissue, hence, we will investigate the validity limits of such effective material models over a wide frequency range (10 MHz - 200 GHz). Embedded in a parameterizable virtual workbench, we implemented a numerical homogenization method using a hierarchical multiscale approach to capture both the dispersive and tensorial electromagnetic properties of the tissue, and determined at which frequency this homogenized model deviated from a full-wave electromagnetic reference model within the framework of a Monte Carlo analysis. Simulations were carried out using a generic hypodermal tissue that emulated the morphology of the microstructure. Results showed that the validity limit occurred at surprisingly low frequencies and thus contradicted the traditional usage of homogenized tissue models. The reasons for this are explained in detail and thus it is shown how both the lower "allowed" and upper "forbidden" frequency ranges can be used for frequency-selective classification/identification of specific material and structural properties employing a supervised machine-learning approach. Using the implemented classifier, we developed a method to identify specific frequency bands in the forbidden frequency range to optimize the reliability of material classification.**

## I. Introduction

CONTACTLESS, non-invasive material characterization/classification based on reflectometry at mm-wave up to THz frequencies is currently gaining a great deal of interest due its potential to resolve both material and structural properties together with the increasing availability of mobile integrated electronic systems in these particular frequency ranges [1]. Recent progress in mm-wave/THz reflectometry has led e.g. to high-precision bulk material characterization of various dielectrics in the mm-wave range of $200-250\,\mathrm{GHz}$ using e.g. a model-based calibration scheme [2], [3]. Broadband THz time-domain spectroscopy (THz-TDS) systems operating in reflection mode in the range of $60\,\mathrm{GHz}$ up to $4\,\mathrm{THz}$ have been successfully employed to e.g. analyze delamination in glass fiber-reinforced composite [4], [5], or in food inspection to reliably distinguish transgenic from non-transgenic rice seeds using machine learning-based classification of spectral fingerprints [6], [7]. An important field of application, which has been thoroughly investigated in the recent years, is the extensive use of mm-wave/THz reflectometry for non-invasive tissue diagnostics [8], [9]. Most of the described approaches retrieve dielectric spectroscopy data using calibration schemes based on homogenized, frequency-dependent tissue models. Such model-based refinement becomes increasingly important when the proper structure of complex tissue morphologies is under consideration [10]. A comprehensive review on the current progress of e.g. mm-wave/THz-based diagnosis of tumorous tissue and neoplasms (beyond the abnormal tissue's water signature) is given in [11]. The biological tissue can be thus viewed as a bona fide benchmark problem for reflectometry-based material characterization especially at mm-wave-/THz frequencies with respect to the structural complexity at different length scales, the large variety of its constituents with associated losses and material contrasts and the inherent anisotropy at the cell level up to the layer structure.

In the mentioned mm-wave/THz reflectometry scenarios, the role of a holistic computational electromagnetics (EM) multiscale tissue model is therefore becoming increasingly important in support of a maximally available sensitivity and selectivity that is achieved by machine learning and regression analysis approaches. The major challenge in developing such tissue model concerns the complex multiscale morphology of the skin, which determines its macroscopic EM properties. Most of the current EM skin models follow a heuristic representation of the skin topology as a multilayer structure as proposed by Alekseev et al. [12] containing typically 3 to 4 homogenized dispersive layers. The material parameters of the latter are retrieved either from fitting models to experimental data [12] or in the framework of the effective medium theory (EMT) using extended analytic mixing formulas with associated multipole Debye models to account for the corresponding frequency dependence. A more rigorous approach builds upon a hierarchically organized multiscale EM model that is rooted in the skin's proper cellular structure, in conjunction with a numerical homogenization procedure of the tissue's microstructure aiming at both the dispersive and tensorial EM material properties. Such multiscale approach has been pioneered by Huclova et al. [13] for human tissue analysis up to 1GHz including the full skin layer morphology together with macroscopic textures like, e.g., the upper and deeper vessel plexus [14] to determine sensitivity and specificity of changes in skin components [15]. An extension to this model up to $1\,\mathrm{THz}$

*Funded by the Deutsche Forschungsgemeinschaft (DFG, German Research Foundation) – Project-ID 287022738 – TRR 196 (Project M03)

[1]General and Theoretical Electrical Engineering (ATE), Faculty of Engineering, and CENIDE – Center for Nanointegration Duisburg-Essen, University of Duisburg-Essen, D-47048 Duisburg, Germany

[2]Corresponding author: Kevin Jerbic (kevin.jerbic@uni-due.de).

has been provided by Saviz et al. [16] using classical mixing formulas for the homogenization of the various tissue layers, and was later complemented by Spathmann et al. [17] to include macroscopic features such as hair follicles and skin wrinkles for frequencies in the range of 100 GHz – 10 THz.

In this paper, we investigate the validity limits of effective material properties of multiphase composites in the context of an EM tissue model. In particular, we discuss the frequency ranges, where commonly used homogenized models for these tissue composites based on the effective material theory start deviating from corresponding hierarchical multiscale full-wave EM reference models. We explicitly focus on hypodermal tissues (HYP) as a generic representation of the tissue composite because of its suitably parametrizable morphology, the low absorption and high material contrast among intra- and extracellular constituents. The multiscale modeling of the HYP tissue starts at the smallest length scale with the microstructure of the randomized cell arrangement confined to a computational supercell with dimensions based on the corresponding correlation lengths in order to grasp the stochastic properties of the underlying composite structure. Our study consists of a systematic two-dimensional (2D) computational EM analysis of HYP tissue surfaces within a generic reflectometry setup for operating frequencies ranging from 10 MHz to 200 GHz. The simulation encompasses 2780 random realizations of HYP tissues for varying structural parameters thus providing a comprehensive Monte-Carlo analysis of the reflectometry scenario. It is worth mentioning that this 2D showcase was only chosen to keep the numerical study manageable within given computer resources. Yet an extension to three dimensions (3D) is straightforward where the findings of the 2D study are transferable to 3D as will be shown in Sec. VII. The results revealed that the validity limits of homogenized tissue models (EMT) compared to a full-wave analysis of the corresponding cell composites appear at astonishingly low frequencies typically in the low mm-wave range [18]. This is in contrast to the traditional use of homogenized layer models in tissue analysis/diagnostics [12], [16] and forces any hierarchically organized multiscale model topology to become strongly tied to a corresponding operating frequency bandwidth.

The remainder of the paper is organized as follows: after an introduction into the methodology and implementation of our multiscale EM simulation workbench for the skin reflectometry scenario in Sec. II, we present a comprehensive Monte-Carlo analysis in Sec. III for the validity limits of various homogenized HYP tissues, given by the operating frequencies from which any effective material representation breaks down. In Sec. IV we show that both, the lower "allowed" frequency band and the upper "forbidden" frequency range (where the EMT loses its validity) can be equally exploited for material characterization/identification. Using a measure based on the local behavior of the difference in the electric and magnetic energy density we can prove that in particular the spectral fingerprints of the tissues in the forbidden range are apt for predicting features of the tissue's microstructure such as the expected values of cell sizes. The subsequent tissue characterization/classification is presented in Sec. V using a machine learning approach based on artificial neural networks (ANN). In addition, ANNs can also be utilized to identify specific bands in the forbidden frequency range which allows to optimize the reliability of the material classification, as demonstrated in Sec. VI. Sec. VII demonstrates the transferability of the research results to three dimensions through corresponding 3D simulations. Sec. VII concludes the study by summarizing the main findings and gives an overview of upcoming research.

## II. Simulation Setup and Methodology

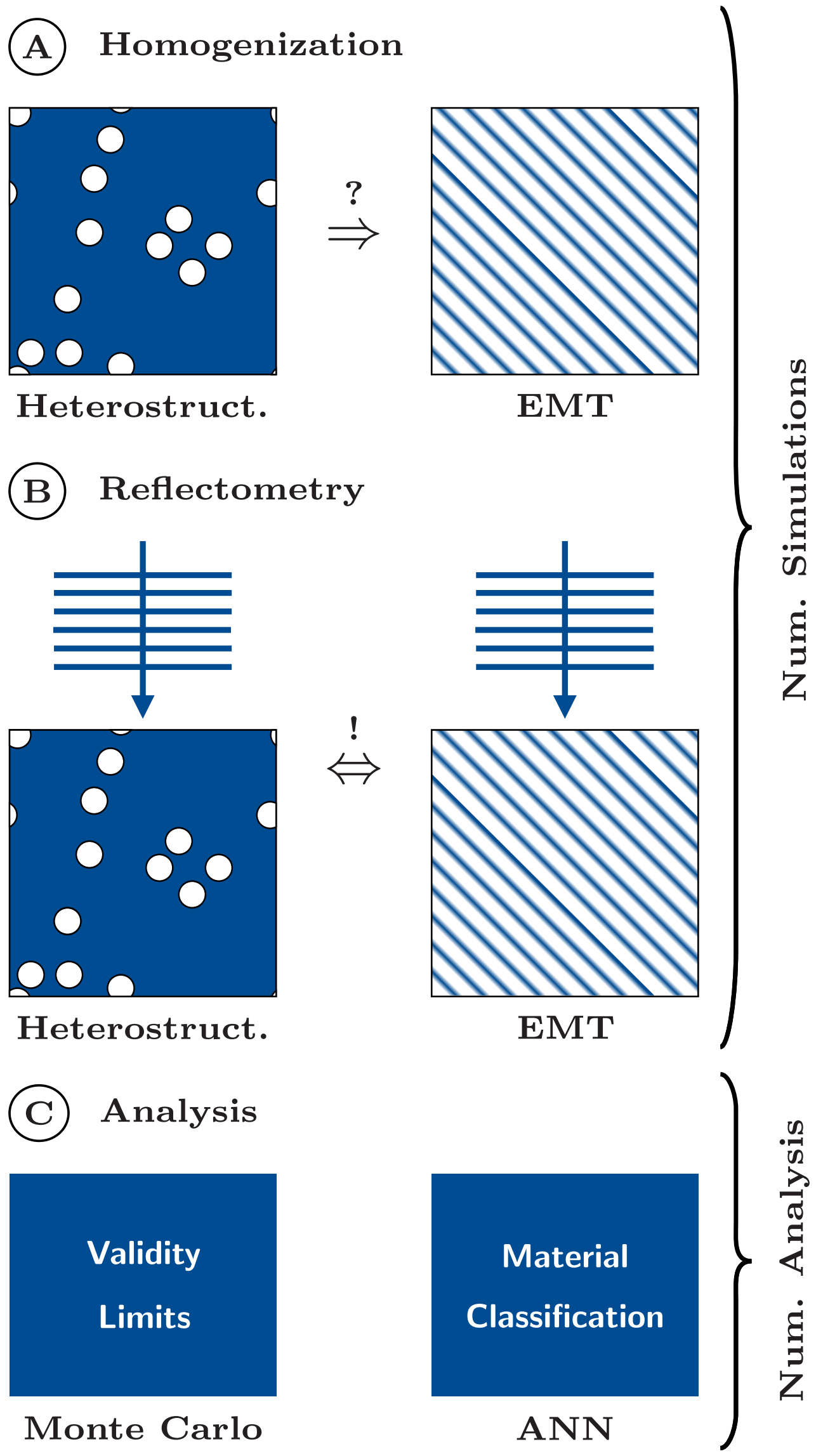

Fig. 1. Pictogram/Schematic of the applied methodology. Abbreviations: effective material theory (EMT); artificial neural network (ANN).

The numerical study of material composites presented here has two aims. First, to reliably explore the validity limits of their homogenized material representation based on the EMT using statistical measures retrieved from a corresponding Monte-Carlo analysis. This implies the numerical analysis of a large number of appropriately parameterized, statistically independent implementations of a given composite. Second, to show, that the validity limit can be exploited to characterize and classify these composites looking at signatures in the frequency response of reflected EM waves, which can be attributed to the underlying morphology of the material structures. To achieve this, a three-stage methodology is applied and implemented in the framework of a multiscale EM simulation workbench whose workflow is shown in Fig. 1.

The methodology starts with (A) an initial homogenization step of the heterogeneous composite using a quasi-static computational EM analysis of a representative composite volume (i.e. the microstructure) yielding the corresponding anisotropic, dispersive EMT representation of the composite. In the next step labelled as (B) the macroscopic effective material properties are then introduced into a computational EM model of generic reflectometry setup. The focus of this step is on the comparison of the frequency-dependent backscattering from the effective material surface with comparable data from the underlying heterogeneous composite when irradiated by an EM plane wave. A validity limit is then defined as the frequency from which a threshold value in the deviation of the two frequency responses (of e.g. $2\,\%$) is exceeded. The last step labelled as (C) encompasses a comprehensive numerical analysis of reflectometry data from large sets of different stochastically generated composites. This type of Monte-Carlo analysis yields reliable statistical measures regarding the validity limits of the underlying effective material representations and is used as the foundation of our material classification with ANNs.

### A. *Randomized representation of the composite microstructure*

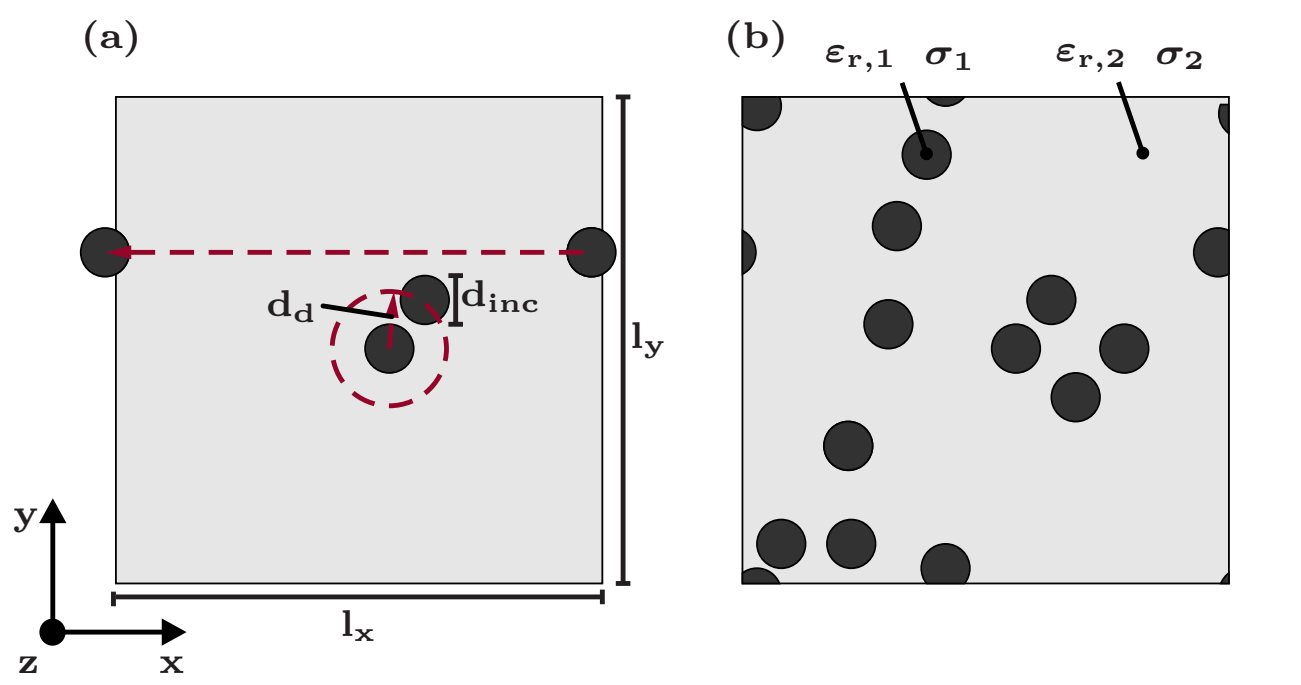

Fig. 2. RSA algorithm for the generation of heterogeneous material structures: (a) overview of the adjustable parameters; (b) example for a generic parameter setup.

In our study we focused on a 2D analysis of HYP as a generic representation of a complex tissue composite. This simplified 2D approach is chosen in order to enable the extensive Monte-Carlo analysis, where an extension to three dimensions (3D) is straightforward as discussed in Sec. VII. The 2D representation of the HYP microstructure is depicted in Fig. 2 as a small computational supercell having monodisperse spherical inclusions (i.e. adipose cells) with $d_{\text{inc}} = 50\,\mu\text{m}$ , $\varepsilon_{\text{r},1} = 80$, and $\sigma_1 = 0.53\,\frac{\text{S}}{\text{m}}$, that are embedded in a homogeneous extracellular medium with $\varepsilon_{\text{r},2} = 50$, and $\sigma_2 = 0.12\,\frac{\text{S}}{\text{m}}$ [19].

The dimensions of the computational supercell are $l_\text{x} = 1\,\text{mm}$ and $l_\text{y} = 1\,\text{mm}$ and chosen accordingly to cover the smallest representative set of inclusions for a given volume fraction $c_\text{v} = \frac{V_\text{inc}}{V_\text{supercell}}$. In order to mimic a realistic HYP microstructure a stochastic representation of the supercell is chosen where the placement of the inclusions is automatically performed by the random sequential addition (RSA) algorithm [20] for a given volume fraction $c_\text{v}$ and a distance parameter $d_\text{d}$ to omit touching and merging of inclusions. The algorithm is modified to enable the periodic continuation of the supercell according to corresponding periodic boundary conditions as illustrated in Fig. 2(a), which reduces the later numerical backscattering analysis to a single supercell.

### B. *Homogenization*

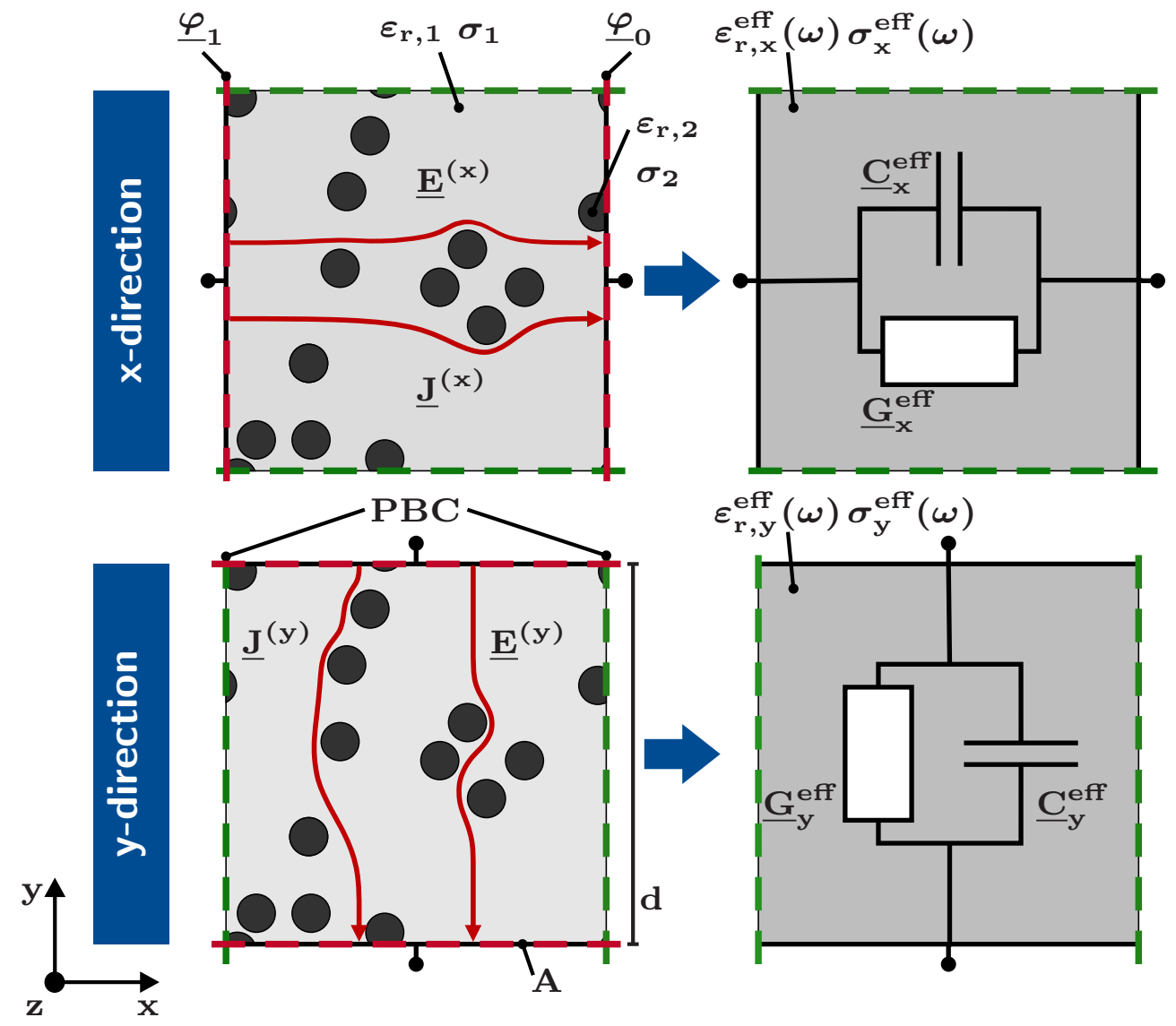

Fig. 3. Schematic of the homogenization procedure: For the tensorial acquisition of the material parameters, the effective material properties are determined successively with respect to the individual spatial directions.

For the homogenization, the 2D supercell containing the heterogenous 2D microstructure of the tissue composite [cf. Fig. 2(b)] is placed into an idealized parallel plate capacitor setup. As depicted in Fig. 3, a time-harmonic voltage with constant amplitude $\hat{\underline{u}} = \underline{\varphi}_1 - \underline{\varphi}_2$ is applied between two opposing supercell edges that are designed as electrodes (i.e Dirichlet boundary conditions).

The two remaining edges are defined as periodic boundary conditions (PBC) in order to suppress fringing fields and to reduce the memory resources of the subsequent quasi-static EM simulation. This approach, which has been implemented with different boundary conditions, has been used in the past to investigate and quantify the influence of interfacial polarization (Maxwell-Wagner) in composite materials compared to calculations based on classical mixing rules [21]. In order to cope with the anisotropy of the underlying microstructure the supercell is then excited in the other (orthogonal) direction while interchanging electrodes with the PBCs (and vice versa) as shown in Fig. 3. The capacitor setup is implemented into the finite-element-method-based (FEM) software package COMSOL Multiphysic [22]. Each 2D simulation model can therefore be interpreted as an infinitely extended 3D capacitor of finite thickness due to two extensions, namely in the direction of the periodic continuation and along the translationally invariant direction normal to the 2D plane. From the time-harmonic quasi-static EM analysis of the supercell respectively the capacitor setup an effective admittance is retrieved that is represented by the equivalent electrical parallel circuit consisting of the elements $G^{eff}$ and $C^{eff}$.

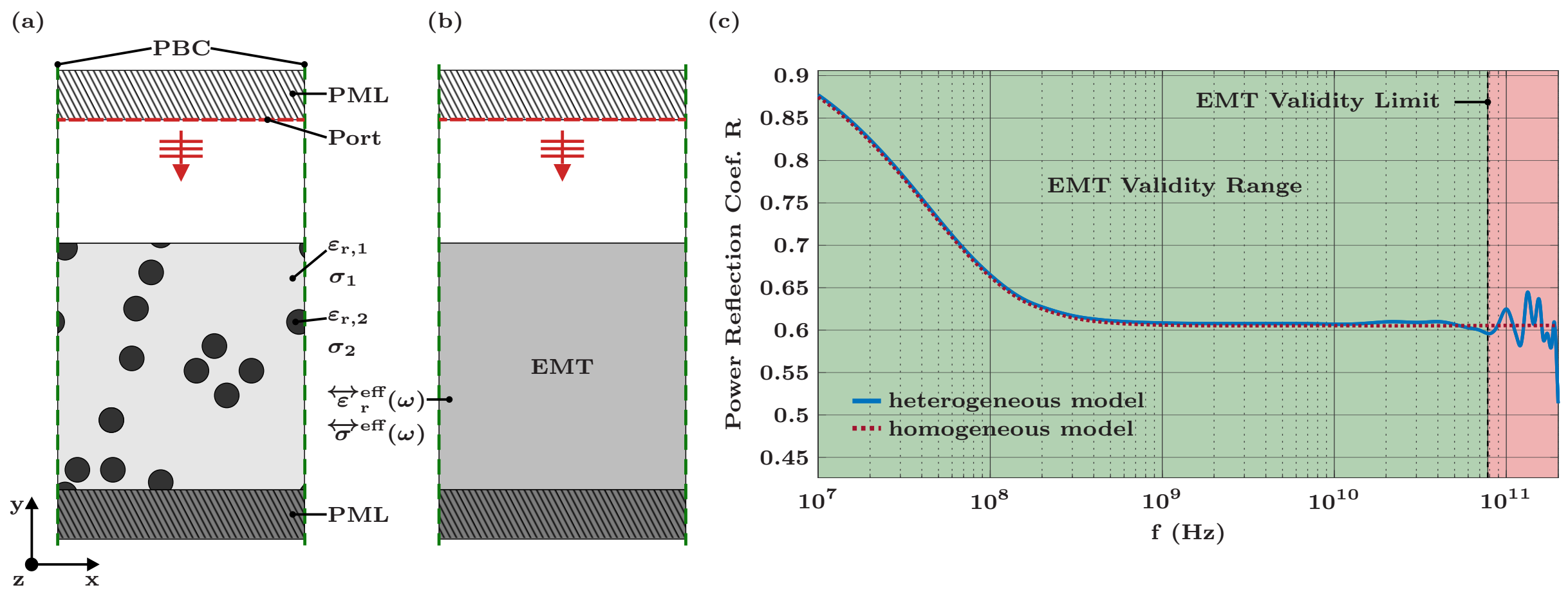

Fig. 4. Comparison between the heterogeneous and homogenized (EMT) model of a generic HYP model: (a) heterogeneous simulation setup; (b) homogeneous simulation setup; (c) spectral responses of the reflectance (showing the typical Maxwell-Wagner roll-off in the MHz range) for $\varepsilon_{r,1} = 80$, $\sigma_1 = 0.53\,\frac{S}{m}$, $\varepsilon_{r,2} = 50$, $\sigma_2 = 0.12\,\frac{S}{m}$, $d_{inc} = 50\,\mu$m, and $c_v = 0.452$.

$$\underline{Y}^{eff}(\omega) = \frac{\hat{\underline{i}}(\omega)}{\hat{\underline{u}}} = G^{eff}(\omega) + j\omega C^{eff}(\omega). \tag{1}$$

as the applied voltage $\hat{\underline{u}}$ and the resulting current $\hat{\underline{i}}(\omega)$ are directly accessible via COMSOL Multiphysics. The effective material properties $\varepsilon^{eff}$ and $\sigma^{eff}$ are thus easily deduced according to

$$\underline{Y}^{eff}(\omega)\frac{d}{A} = \frac{\hat{\underline{i}}(\omega)}{\hat{\underline{u}}} \cdot \frac{d}{A} = \underbrace{\sigma^{eff}(\omega) + j\omega\varepsilon_0\varepsilon_r^{eff}(\omega)}_{\underline{\sigma}^{eff}(\omega)}. \tag{2}$$

where $d$ is the parallel plate distance and $A$ stands for area of the electrode. In Eq. 2 the righthand term can be interpreted as the complex effective conductivity $\underline{\sigma}^{eff}(\omega)$ of the homogenized effective medium from which the required material parameters directly follow

$$\sigma^{eff}(\omega) = \Re\{\underline{\sigma}^{eff}\} \tag{3}$$

and

$$\varepsilon_r^{eff}(\omega) = \frac{\Im\{\underline{\sigma}^{eff}\}}{\omega\varepsilon_0}. \tag{4}$$

In order to consider anisotropies in the effective medium the quasi-static capacitor analysis is performed with excitation in both x- and y-directions (cf. Fig. 3) yielding corresponding frequency-dependent second-rank tensors for the effective permittivity and effective conductivity, respectively.

For the conductivity

$$\overleftrightarrow{\sigma}^{eff}(\omega) = \begin{pmatrix} \sigma_x^{eff}(\omega) & 0 \\ 0 & \sigma_y^{eff}(\omega) \end{pmatrix} \tag{5}$$

and for the permittivity

$$\overleftrightarrow{\varepsilon}_r^{eff}(\omega) = \begin{pmatrix} \varepsilon_{r,x}^{eff}(\omega) & 0 \\ 0 & \varepsilon_{r,y}^{eff}(\omega) \end{pmatrix}. \tag{6}$$

These are the homogenized, frequency-dependent effective material representations that are later introduced (in accordance with the multiscale-modeling approach) into the simulation model of the generic reflectometry setup.

### C. *Reflectometry*

The generic reflectometry setup is used to analyze the reflection of an impinging EM plane wave from a surface system that is defined by a plane boundary between air and either a heterogeneous composite material or its homogenized EMT representation (cf. Fig. 4). The plane wave is excited from a non-reflective port under the angle of incidence $\alpha_{inc}$ (here $0\,$deg) having either s- or p-polarization with respect to the reflection plane (i.e. x-y plane). The amplitude of the incident plane wave is assigned to a constant input power $P_0$ whereas the reflected power $P_R$ is detected by the same non-reflective port from which the desired measure, namely the power reflection coefficient $R = \frac{P_R}{P_0}$ is calculated. The absorption coefficient $A = \frac{P_A}{P_0}$ is determined by a corresponding volume integration of the material losses $P_A$ within the material structure. Top and bottom of the computational domain are terminated with perfectly matched layers (PML), which are assigned to the effective material parameters defined by $\overleftrightarrow{\varepsilon}_r^{\text{eff}}(\omega)$ and $\overleftrightarrow{\sigma}^{\text{eff}}(\omega)$ and to the material parameters of the air respectively. Similar to the capacitor setup, PBCs are also introduced into the reflectometry model to omit fringing fields while extending the randomized super cell to a periodic representation of the composite surface layer.

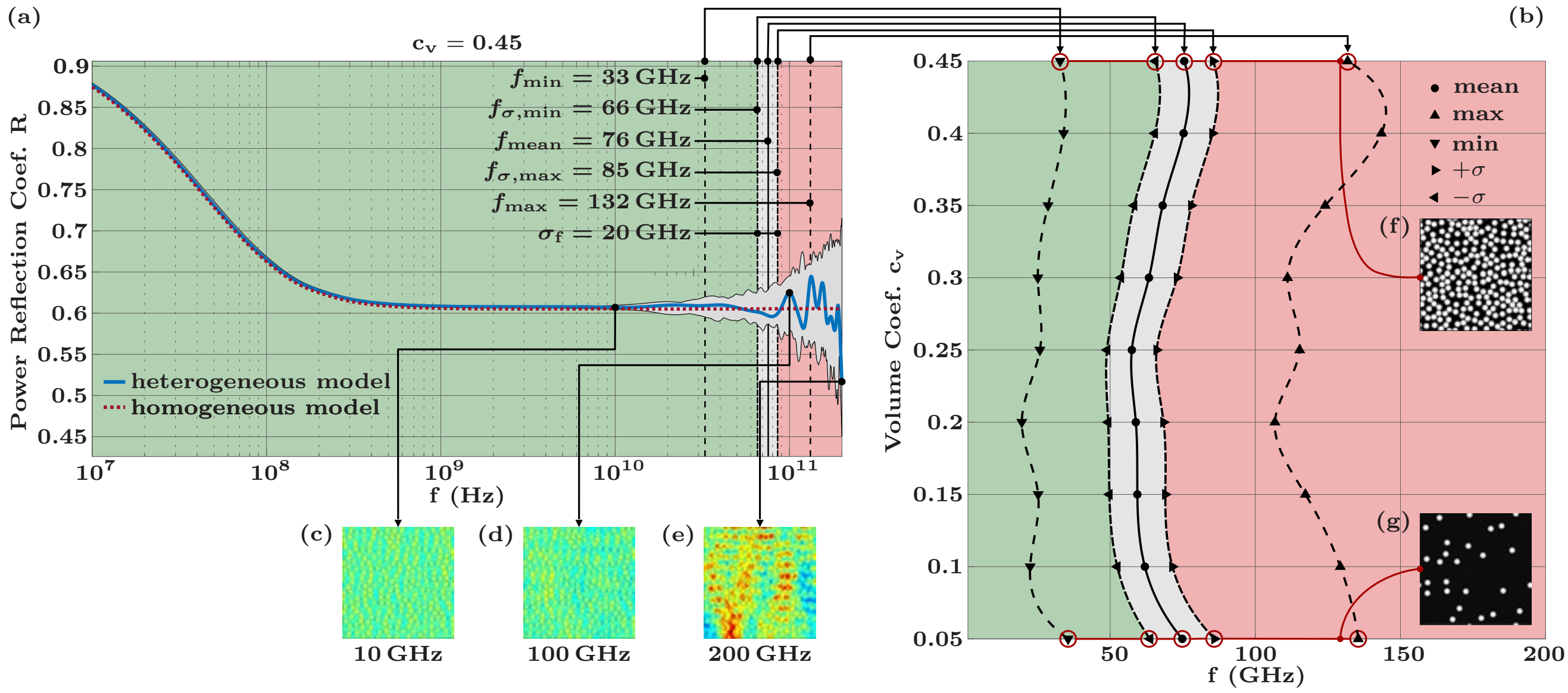

Fig. 5. Validity limits for several generic HYP derivatives: (a) spectral responses of the reflectance (showing the typical Maxwell-Wagner roll-off in the MHz range) of 220 implementations for $\varepsilon_{r,1} = 80$; $\sigma_1 = 0.53\,\mathrm{S/m}$; $\varepsilon_{r,2} = 50$; $\sigma_2 = 0.12\,\mathrm{S/m}$; $d_{inc} = 50\,\mu\mathrm{m}$; and volume fraction $c_v = 0.45$; (b) Validity limits of the derivatives of a heterogeneous material structure (here the HYP tissue); (c), (d), (e) examples of the electric field distribution $|\vec{E}|$ at various frequencies (i.e. at 10 GHz, 100 GHz, 200 GHz, respectively); (f), (g) examples of the analyzed microstructures (i.e. for $c_v = 0.45$ and $c_v = 0.05$ respectively). The validity range of the EMT material model is colored in «green» while the forbidden range is marked «red».

A first example for analyzing the breakdown of the effective material representation is given in Fig. 4(c) for a HYP tissue with a cell volume fraction of $c_v = 0.452$. The simulation has been performed on PC equipped with an Intel i7-6700k processor (4 cores) and 64 GB DDR4 RAM. In the given frequency interval, 260 frequency points have been simulated with a MUMPS solver whereby the density of the frequency points for higher frequencies has increased. The simulation for each frequency point lasts 21 s and includes the simulation of heterogeneous microstructure and its homogenized representative in p- and s-polarization. The blue curve represents the power reflection coefficient of the heterogeneous composite model and the dotted red curve that of the homogeneous EMT model, for p-polarized excitation in a frequency range from 10 MHz to 200 GHz. Since in s-polarization the electric field points in the translation invariant direction of the composite and therefore homogenization is not defined, the comparison between the reflectance of the heterogeneous material structure and its homogenized representative is only done for p-polarization. The validity limit of the EMT representation for a deviation of 2 % between the two frequency responses is located at $f_{val} = 78$ GHz for this particular HYP tissue realization. The validity range of the homogenized EMT representation therefore extends below the validity limit, whereas the "forbidden" range above this limit is characterized by strong variations in the frequency response associated to the heterogenous tissue model which defines a characteristic fingerprint apt to be exploited for material classification.

## III. Monte Carlo Analysis of Validity Limits

Since the analysis of a single HYP tissue structure has only minor significance for a reliable determination of the associated EMT representation's validity limit, the procedure presented in Sec. II is thus performed for a large number of randomly generated realizations of the same HYP tissue based on a corresponding structural parameter set labelled as $P_i$. Each set $P_i$ contains fixed values for both the minimal separation distance $d_d = 1.05$ and the diameter $d_{inc} = 0.05\,\mathrm{mm}$ of the HYP's spherical inclusions together with a specific volume fraction $c_{v,i} \in \{0.05, 0.1, \dots, 0.45\}$. In this sense 9 different HYP tissue types are addressed according to the parameter sets $P_1, \dots, P_9$ with altered volume fractions $c_{v,1}, \dots, c_{v,9}$. Within the framework of a Monte-Carlo analysis 220 statistically independent microstructures with randomly distributed inclusions were created for each $P_i$ respective $c_{v,i}$ which are considered in the following as realizations of the parameter set $P_i$. Examples of such microstructures are shown in Fig. 5(f) and 5(g) for the parameter sets $P_9 := (d_d = 1.05,\ d_{inc} = 0.05,\ c_v = 0.45)$ and $P_1 := (d_d = 1.05,\ d_{inc} = 0.05,\ c_v = 0.05)$ respectively. As an exemplary case the statistical analysis for the HYP sample $\mathrm{P}_9$ ($c_{\mathrm{v}} = 0.45$) with 220 randomly generated realizations of the heterogeneous composite is illustrated in Fig 5(a). The frequency responses of the simulated power reflection coefficient of all realizations confine to the gray shaded area between the top and bottom envelope, with the blue curve labeling a single representation of this spectral set which has already been depicted in Fig. 4(c). The red-dotted curve shows the performance of the HYP tissues' homogenized EMT model. The validity limits of the EMT homogenization are determined as the frequency from where the power reflection of the homogeneous EMT model and the heterogeneous composite model deviate by more than $2\,\%$ yielding thus for e.g. $\mathrm{P}_9$ a corresponding dataset with 220 values with a mean $f_{\mathrm{mean}} = 76\,\mathrm{GHz}$ which in the following is defined as the representative value for the validity limit. The associated standard deviation is $\sigma_{\mathrm{f}} = 20\,\mathrm{GHz}$ leading to a first upper and lower bound of $f_{\sigma,\mathrm{min}} = 66\,\mathrm{GHz}$ and $f_{\sigma,\mathrm{max}} = 86\,\mathrm{GHz}$ whereas an absolute minimum and maximum of this limit amounts to $f_{\mathrm{min}} = 33\,\mathrm{GHz}$ and $f_{\mathrm{max}} = 132\,\mathrm{GHz}$ as depicted in Fig. 5(a). The validity range of the EMT material model is colored in «green» while the forbidden range, above which the EMT representation breaks down is marked «red». The overall Monte-Carlo analysis has in the following been carried out for all parameter sets $\mathrm{P}_1, \ldots, \mathrm{P}_9$ respective volume fractions $c_{\mathrm{v},1}, \ldots, c_{\mathrm{v},9}$ (cf. Fig. 5(b)) and revealed validity limits for HYP tissues at astonishingly low frequencies around $60 - 80\,\mathrm{GHz}$. This is an interesting result as it may challenge assumptions made in commonly used tissue models for mm-wave [12] and THz frequencies [11]. In addition, the statistical evaluation of all parameter sets has also led to validity limits with only weak dependence against the volume fraction of the lipid droplets (i.e. inclusions) in the adipose HYP tissue. Fig. 5(c)-(e) depicts the electric field distributions $|\vec{E}|$ within the microstructure of the heterogeneous composite for three distinct frequencies in the blue curve. As expected from EMT are these field profiles are quite homogeneous at frequencies (e.g. $10\,\mathrm{GHz}$) in the validity range but display interference patterns at «forbidden» frequencies (e.g. $100\,\mathrm{GHz}$ and $200\,\mathrm{GHz}$) with emerging spatial correlations to the underlying microstructure. In the context of reflectometry-based tissue characterization these correlations will establish the possibility of using the spectral fingerprints in the «forbidden» range as shown by the blue curve in Fig. 5(a) to classify microscopic features of the heterogeneous microstructures, whereas the well-behaved frequency response in the validity range allows for the extraction of macroscopic quantities such as the effective material parameters and, for example, the underlying volume fraction. This will be discussed in the next section.

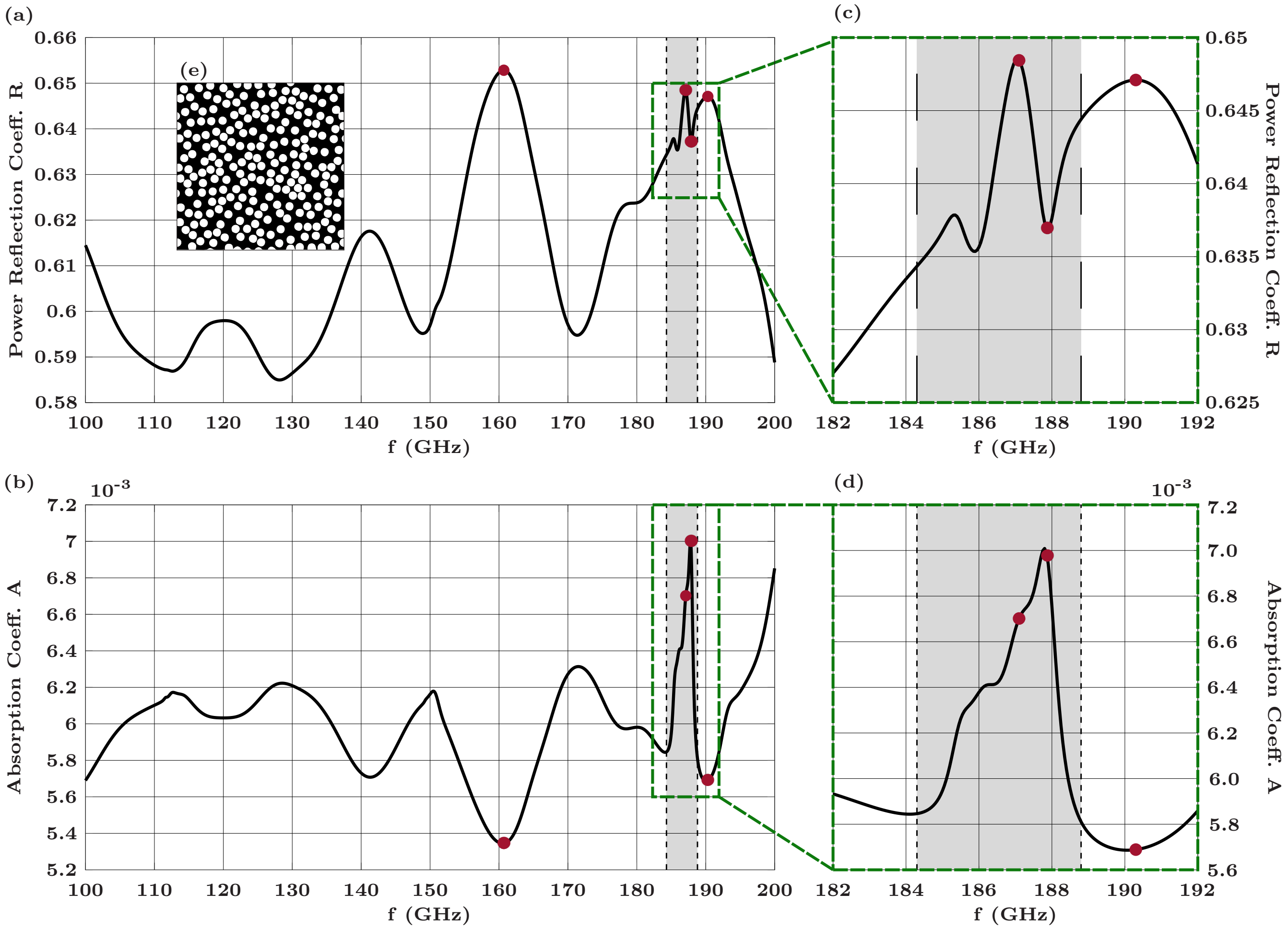

Fig. 6. Spectral response in the volatile frequency range between 100 GHz and 200 GHz: (a) reflectance; (b) absorptance; (c), (d) detailed enlargements of the reflectance and absorptance in the subinterval between 182 GHz and 188 GHz; (e) example of the analysed HYP realization (i.e. for $\varepsilon_{\mathrm{r},1} = 80$; $\sigma_1 = 0.53\,\mathrm{S/m}$; $\varepsilon_{\mathrm{r},2} = 50$; $\sigma_2 = 0.12\,\mathrm{S/m}$; $d_{\mathrm{inc}} = 50\,\mu\mathrm{m}$; and volume fraction $c_{\mathrm{v}} = 0.45$).

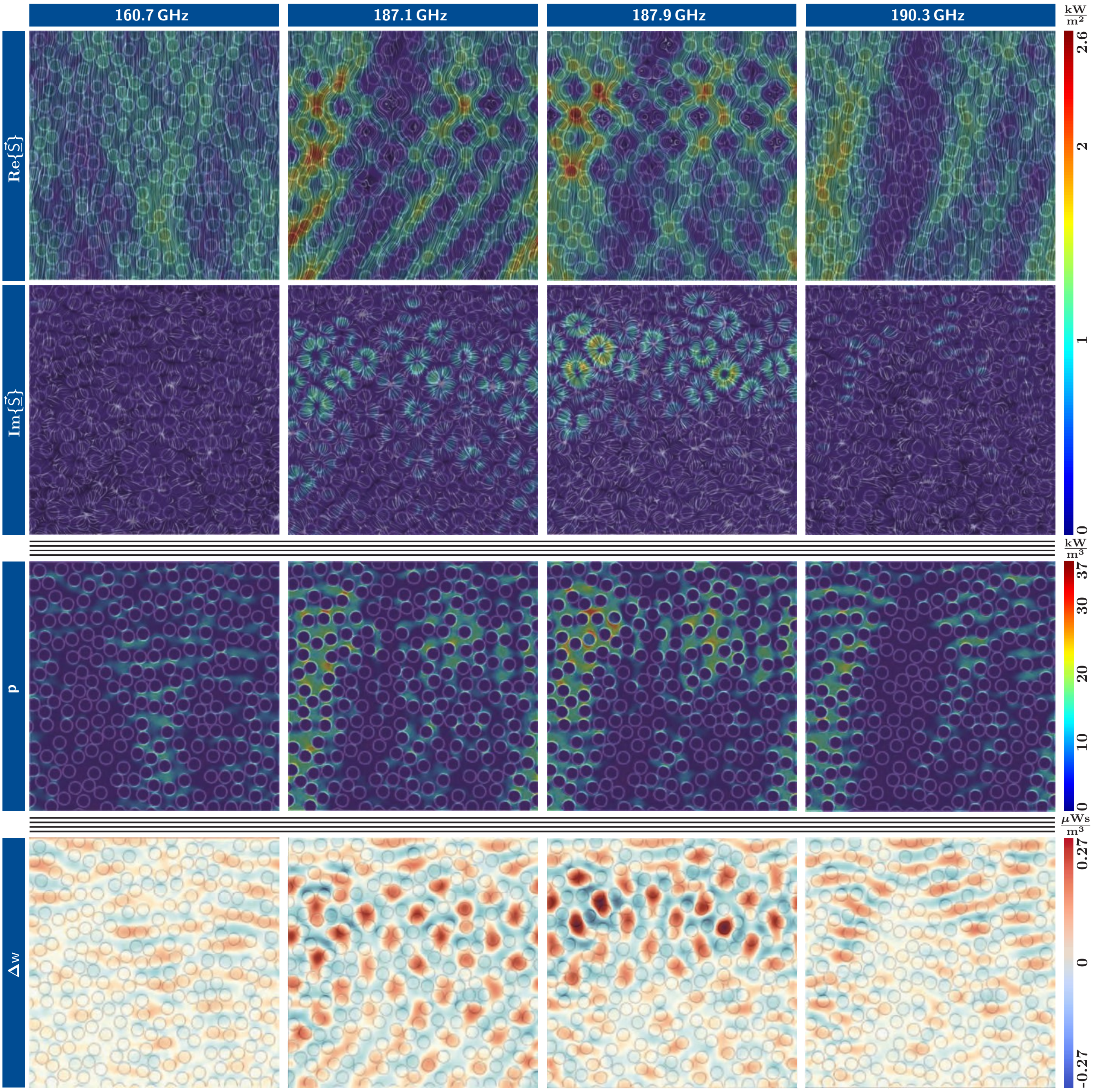

Fig. 7. Graphical evaluation of the EM fields within the micro structure for the frequency points 160.7 GHz, 187.1 GHz, 187.9 GHz and 190.3 GHz (i.e. for $\varepsilon_{\mathrm{r},1} = 80$; $\sigma_1 = 0.53\,\mathrm{S/m}$; $\varepsilon_{\mathrm{r},2} = 50$; $\sigma_2 = 0.12\,\mathrm{S/m}$; $d_{\mathrm{inc}} = 50\,\mu\mathrm{m}$; and volume fraction $c_{\mathrm{v}} = 0.45$). From the top to the bottom row: the real part of the Poynting Vector $\mathrm{Re}\{\vec{\underline{S}}\}$ the imaginary part $\mathrm{Im}\{\vec{\underline{S}}\}$ the loss density $p$ and the difference of enegery densities $\Delta w$.

## IV. On the Information Content of the Forbidden Frequency Band

The existence of validity limits has direct consequences on the use of EMT-based multiscale models for multi-layered tissue systems such as skin. In such models the maximum frequency is determined by the tissue layer with the lowest validity limit, yielding thus an allowed operating frequency band which becomes specific for the overall skin model. At higher frequencies outside this band, the described homogenization of the particular tissue layer is not applicable anymore and its further analysis requires an accurate modelling of its proper microstructure, which actually leads to a modified multiscale skin model. An ultra-broadband tissue analysis must therefore consider simultaneous structural changes in the tissue models during the frequency-domain simulation while relying on an appropriately prepared set of multiple model representations. These multiple representations can be set up prior to numerical analysis when e.g. when using machine learning-based predictions of the involved validity limits.

As indicated in the previous section the increasing impact of the microscopic properties on the frequency response beyond the validity limit opens up the possibility to identify and classify specific features of the tissue's microstructures just based on the associated spectral fingerprint in the forbidden frequency range. This reasoning can be underpinned by correlating characteristic features in this spectral fingerprint with emerging interference patterns in the microstructure. Within an illustrative example we studied specific frequency points in the spectral fingerprints of the power reflection $R$ and absorption $A$ between 100 GHz and 200 GHz for a single HYP tissue realization ($c_\mathrm{v} = 0.45$) in conjunction with the associated EM field patterns in the microstructure.

A field quantity of particular significance here is the difference in electric and magnetic energy density $\Delta w$. This measure has already proven successful in the assessment of advanced MRI coils [23]. It is derived from the time-harmonic version of the Poynting theorem, which stays for the conservation of the EM energy flux $\underline{\vec{S}}$ (i.e. the Poynting vector) as shown in Eq. (7) where $\underline{\vec{S}}$ has to fuel areas with dissipation loss and at the same time to compensate for temporal changes in the reactive power.

$$\nabla \cdot \underline{\vec{S}} = \underbrace{-\frac{1}{2}\sigma|\underline{\vec{E}}|^2}_{p} -j2\omega \underbrace{\frac{1}{4}\left(\mu|\underline{\vec{H}}|^2 - \varepsilon_0\varepsilon_\mathrm{r}|\underline{\vec{E}}|^2\right)}_{\Delta w = w_\mathrm{m} - w_\mathrm{e}}, \qquad (7)$$

The term of the reactive power carries the difference $\Delta w$ in the electric and magnetic energy density where $\Delta w$ emerges only if the energy density distributions of both field quantities are spatially separated. The quantity $\Delta w$ thus indicaties a locally confined resonant enhancement of reactive fields in the tissue's very microstructure and aims on the other hand at potential «hot spots» of the loss density $p$, which is mainly due to oscillating balancing currents between these separated energy densities. The dissipated power $P_\mathrm{A}$ in the tissue results from integrating the power loss density $p$ over the tissue volume yielding the absorption $A$. As depicted in Fig. 7, $\Delta w$ provides the most selective map of the resonant loss enhancement which is highly correlated to the tissue's microstructure and therein to the randomized HYP cell distribution. With this connection between microstructure affine field patterns and specific signatures in the spectral fingerprint we reason that the spectral responses in the forbidden frequency range contain enough information for the characterization of the underlying microstructure. In the following the spectral responses of both, power reflection $R$ and absorption $A$ of the given HYP tissue are plotted in Fig. 6 for the mentioned region of interest in the forbidden frequency range. Apart from the expected opposite behavior between the frequency response of $R$ and $A$ [cf. Fig. 6(a) and (b)] where a reflection maximum is accompanied by an absorption minimum, there are distinct subintervals such as between 184 GHz and 188 GHz (gray shaded) with a different behavior. An enlarged view of this specific region is plotted in Fig. 6(c) and 6(d) and shows both, $R$ and $A$, forming unique spectral substructures. In the following we analyzed local extrema at frequencies 160.7 GHz, 187.1 GHz, 187.9 GHz and 190.3 GHz outside and within the sub-interval (labelled by red dots).

The associated field patterns regarding flowing, dissipating and stored energy is depicted in Fig. 7 partly using a generalized field line representation provided by the so-called line integral convolution (LIC) method [24]. Comparing these fields inside the subinterval (at 187.1 GHz and 187.9 GHz) with those outside (at 160.7 GHz and 190.3 GHz) indicates that the imaginary part of the energy flux $\underline{\vec{S}}$ and even more the quantity $\Delta w$ (which is actually derived from $\mathrm{Im}\{\underline{\vec{S}}\}$) yield the most distinct correlations to the underlying microstructure. We can therefore assume that especially these unique spectral substructures in the spectral responses significantly contribute to the information content about the tissue's microstructure. This makes it easy to understand that promising tissue characterization schemes based on skin surface reflectometry could implicitly use such unique spectral features and eventually exploit them in the framework of supervised learning-based classification approach. In the following section we therefore present a successful procedure that utilizes the spectral fingerprints of the HYP structures in both, the valid and the forbidden frequency range for reflectometry-based tissue classification with ANN.

## V. ANN-Based Tissue Classification

### A. Datasets for Microstructure Classification

For the tissue classification based on ANN we have exploited both the forbidden frequency range to predict the microstructures of the HYP tissues (e.g. the expectation values of the cell sizes) and the validity range to classify aggregated HYP tissue properties such as the volume fractions of the adipose cells. In this regard the Monte-Carlo analysis described in Sec. III is extended by 4 additional parameter sets $\tilde{\mathrm{P}}_1$ to $\tilde{\mathrm{P}}_4$ with fixed values for $d_\mathrm{d} = 1.05$ and variable structural parameters for $d_\mathrm{inc} \in \{0.05\,\mathrm{mm}, 0.1\,\mathrm{mm}\}$ and $c_\mathrm{v} \in \{0.2, 0.35\}$. The reflectometry data of these 4 generic HYP tissues is given in Fig. 8, where each of them is numerically represented by 200 statistically independent realizations of the underlying microstructure. To investigate the performance of the classification procedure the analyzed spectral range is subdivided into 2 frequency intervals $\mathrm{F}_\mathrm{val}$ and $\mathrm{F}_\mathrm{for}$ as depicted in Fig. 8(a). The interval $\mathrm{F}_\mathrm{val} := [10\,\mathrm{MHz}; 30\,\mathrm{GHz}]$ covers the characteristic Maxwell-Wagner roll-off associated to relaxation effects of polarization charges at the cell surface, $F_{for} := [30\,\mathrm{GHz}; 200\,\mathrm{GHz}]$ includes the range of the proper validity limit and the forbidden frequency range.

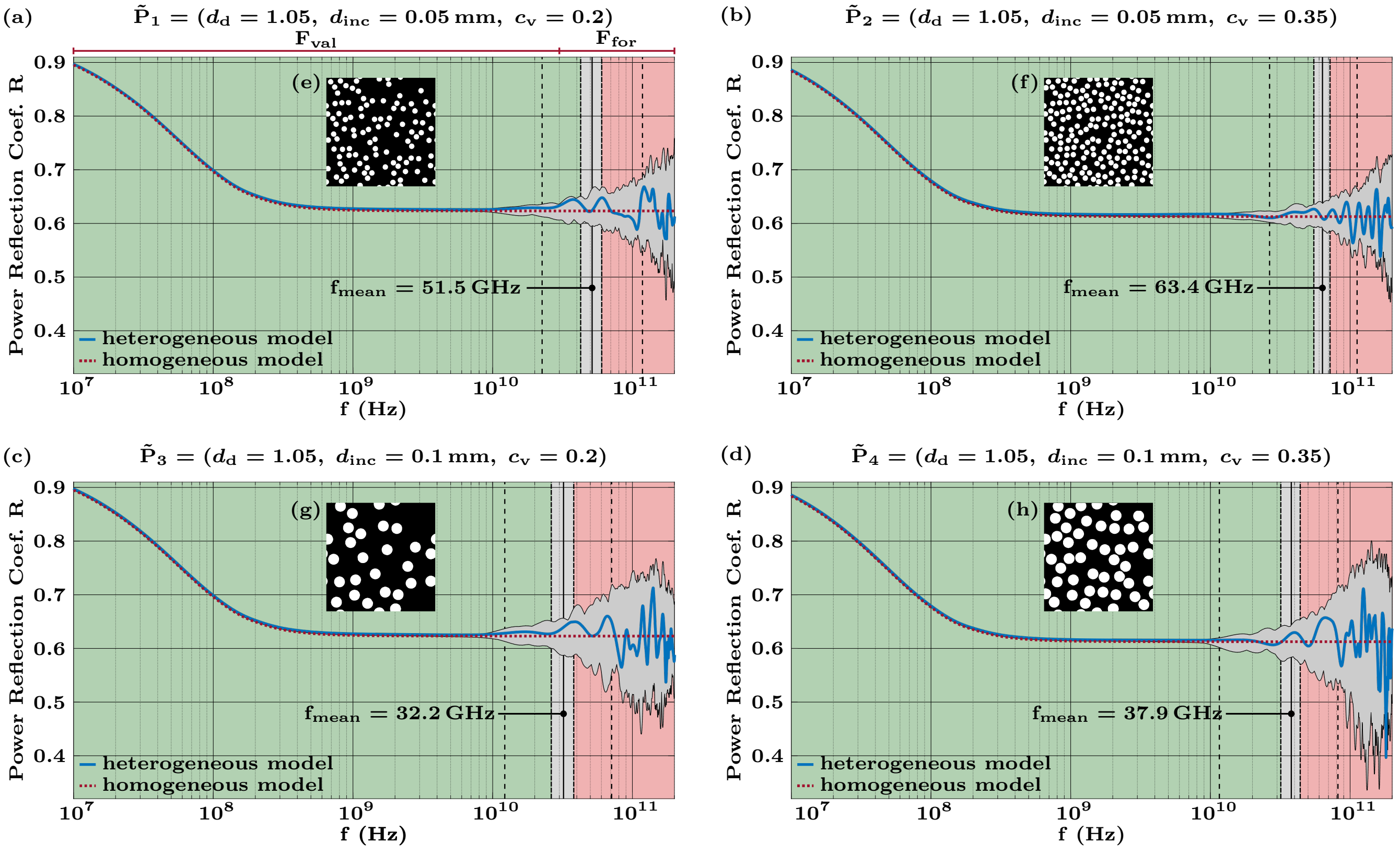

Fig. 8. Overview of the parameter sets $\tilde{P}_1$ to $\tilde{P}_4$: (a)-(d) spectral responses of the reflectance of 200 statistically HYP realizations of the parameter sets $\tilde{P}_1$ to $\tilde{P}_4$ respectively; (e)-(h) examples of the analysed microstructures.

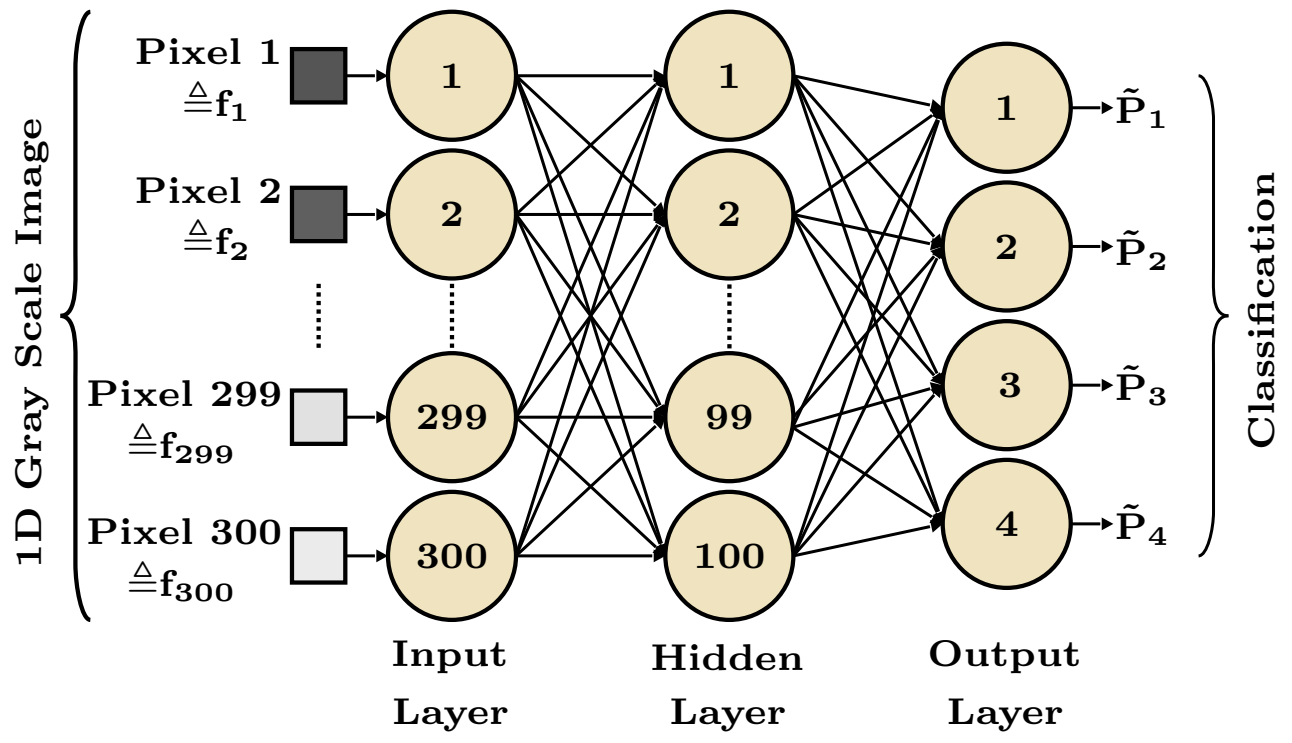

Fig. 9. Overview of the network architecture: A 1D-image input layer for the processing of 300 image pixels consisting of 300 neurons, a subsequent fully connected hidden layer with 100 neurons which has a rectified linear unit (ReLU) activation function followed by a fully-connected output classification layer (using a softmax function) in which the number of neurons is derived from the number of categories to be classified ($\tilde{P}_1$ to $\tilde{P}_4$).

As preparation to the ANN-based classification, the spectral fingerprint of each randomized realization is sampled at 300 discrete frequency points within the complete frequency range as well as in each of the given subintervals $F_{val}$ and $F_{for}$ where the sampled values of the power reflection are displayed according to a grayscale coding. The resulting dataset encompassing the realizations of the corresponding frequency interval can therefore be depicted as a 2D grayscale image with the 300 image pixels along the frequency axis and the corresponding label indices of the involved spectral fingerprints along the other dimension.

The resulting image recognition-based material classification is implemented by the use of shallow ANNs as depicted in Fig. 9 and performed with the image processing library of the MATLAB programming environment (version R2018a). The ANNs consist of a 1D-image input layer for the 300 image pixels consisting of 300 neurons, a subsequent fully connected hidden layer with 100 neurons which has a rectified linear unit (ReLU) activation function followed by a fully-connected output classification layer (using a softmax function) in which the number of neurons is derived from the number of categories to be classified (in this case 4 classes referring to the parameter sets $\tilde{P}_1$ to $\tilde{P}_4$ depicted in Fig. 8). As training scheme, the stochastic gradient descent with momentum (SGDM) method was applied to 80 % of the available data where every network is trained over 3000 epochs yielding an overall training time of approximately 5 minutes. For comparability of the classification results, the network architecture (cf. Fig.9) is used for all frequency

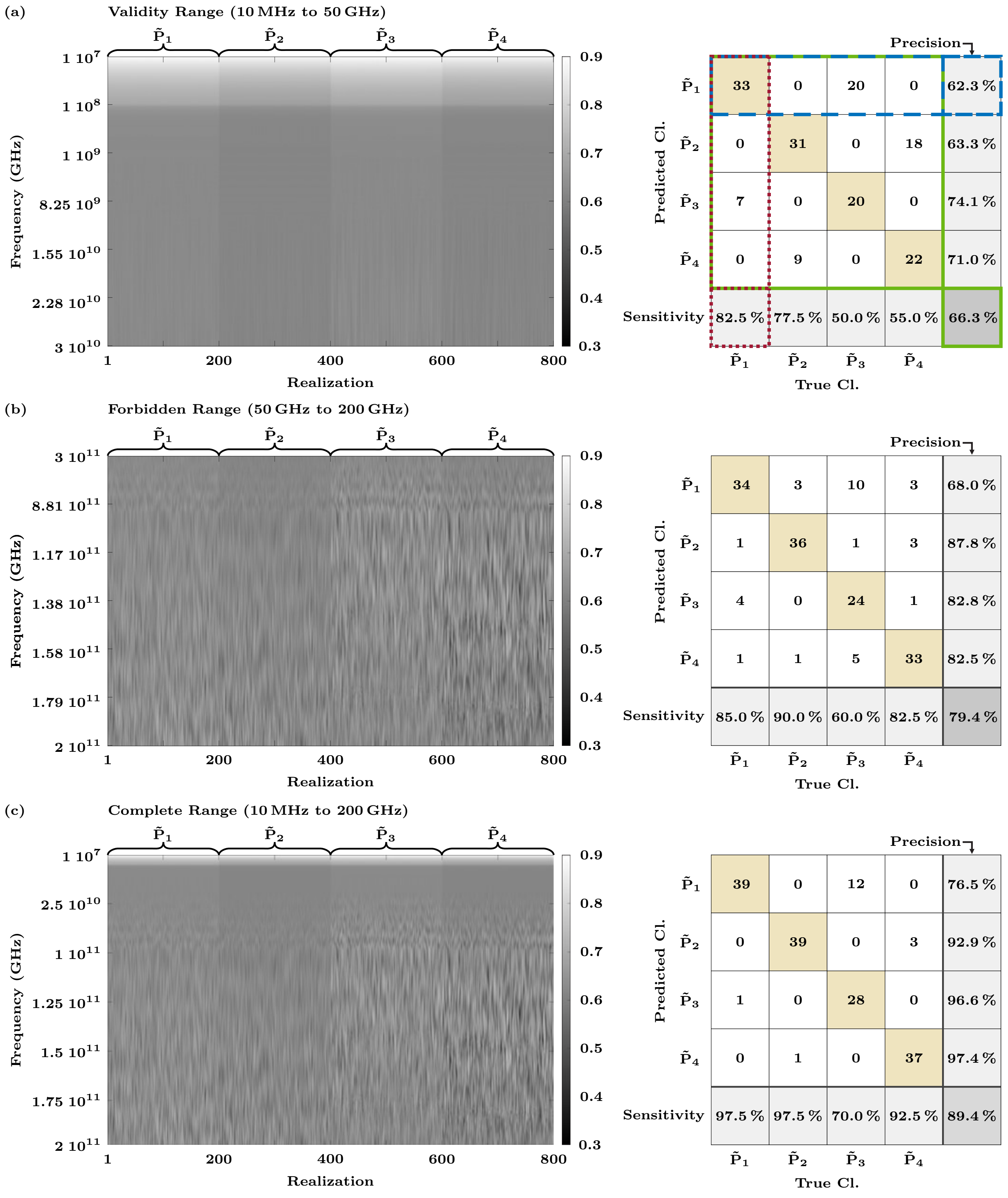

Fig. 10. Overview of the ANN-based classification of 4 generic tissue derivatives represented by the parameter sets $\tilde{P}_1$ to $\tilde{P}_4$ using the spectral fingerprint in the validity, forbidden and complete frequency range summarized in confusion matrices in (a) to (c), respectively. The corresponding grayscale images display the reflection coefficient of each realization of the considered parameter sets sampled at 300 discrete frequency points in a grayscale coding.

intervals. The remaining $20\,\%$ of data is used to test the trained ANNs' performance.

The results of the classification in the validity and the forbidden range as well as in the complete frequency range are depicted in Fig. 10(a) to (c) respectively. In these, the spectral fingerprints of all 800 realizations are plotted according to the mentioned grayscale coding and represent the input data for the ANNs. The results of the ANN-based classifications are summarized in the corresponding confusion matrices (test set data). These confusion matrices compare the classes of the individual realizations predicted by the ANN against their actual classes. The correctly assigned realizations are arranged on the main diagonals of the matrices and the incorrectly assigned ones on the minor diagonals. The overall performance of the classification is evaluated by three metrics, namely *accuracy*, *precision* and *sensitivity*. The *accuracy* (green frame Fig. 10(a)), which can be regarded as the overall quality measure of the classification, is calculated as the ratio of correct predictions to all predictions of a classification. The *precision* (blue frame) is calculated as the ratio of the correct predictions to all predictions in favor of within a particular class. The *sensitivity* (red frame) is calculated as the ratio of the correct predictions to all actual elements of a particular class.

The lowest accuracy is achieved with $66.3\,\%$ in the validity range. If the data from the forbidden range is used, the accuracy improves to $79.4\,\%$. This shows that within the forbidden range there is a high information content for the classification of microstructures. However, if the entire spectrum is used for classification, an accuracy of $89.4\,\%$ can be achieved. This implies that the information content of both ranges is complementary and makes the classification more reliable.

In order to investigate the extent to which the validity and the forbidden range complement each other, we will examine in the following subsection whether specific properties of the microstructure can be classified in specific frequency ranges.

## B. Classification of Specific Material Properties

In this section we will show that the acquisition of the HYP tissues' specific microstructure is possible using the targeted evaluation of the power reflection coefficient used as a spectral fingerprint by ANNs within specific frequency intervals.

For this purpose, we will perform a binary classification with respect to the volume fraction $c_\text{v}$ and the cell diameter $d_\text{inc}$. To implement the binary classification, the output layer of the ANNs used is reduced to two neurons (cf. Fig. 11(a)). In addition to the modification of the ANN's output layer, we reorganized the spectral fingerprints of the parameter sets $\tilde{\text{P}}_1$ to $\tilde{\text{P}}_4$ according to Fig. 11(b). The material parameter indicating the classification target is now considered as the primary parameter and the other as the secondary one acting as a perturbation. Following this principle, parameter sets of the same primary parameters are grouped together into two classes of the same size. As a result, we obtain two classification cases which still include all the 800 realizations.

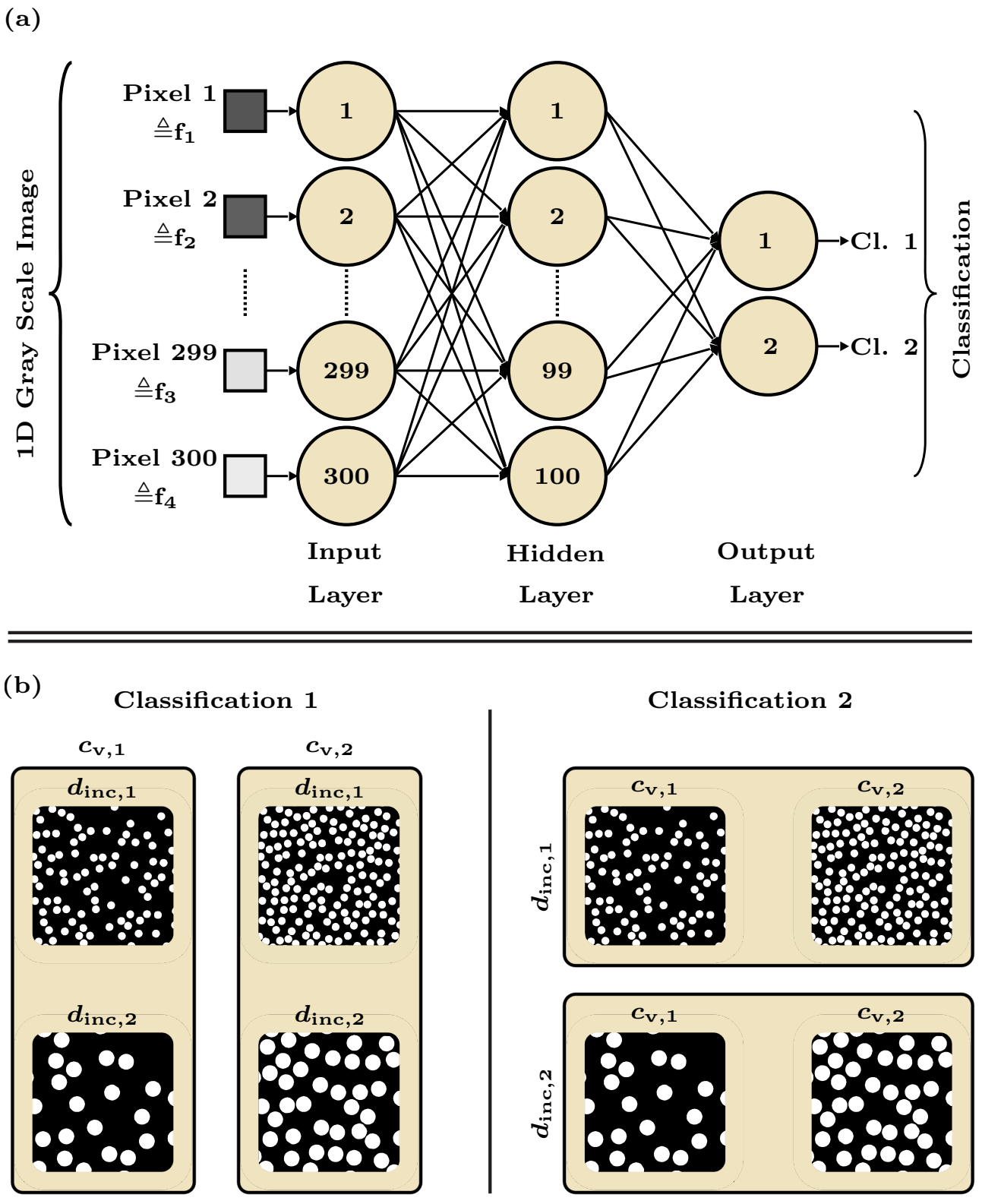

Fig. 11. Implemented modifications for the binary classification: (a) Reduction of the output layer to two neurons. (b) Reorganization of the parameter sets $\tilde{\text{P}}_1$ to $\tilde{\text{P}}_4$ into two classification cases. Classification 1 targeting the volume coefficient with the classes $c_{\text{v},1} = 0.2$ and $c_{\text{v},2} = 0.35$ and classification 2 the inclusion diameter with the classes $d_{\text{inc},1} = 0.05\,\text{mm}$ and $c_{\text{inc},2} = 0.1\,\text{mm}$.

To investigate the frequency specificity of the classification, the analyzed spectral range is rearranged and subdivided into 5 new frequency intervals $\text{F}_1$ to $\text{F}_5$. The interval $\text{F}_1 := [10\,\text{MHz}; 10\,\text{GHz}]$ covers the validity range up to a point where no significant variance of the heterogeneous models' reflection coefficient exists, $\text{F}_2 := [10\,\text{GHz}; 100\,\text{GHz}]$ includes an extended transition range between the validity and forbidden range including the validity limit, whereas $\text{F}_3 := [100\,\text{GHz}; 150\,\text{GHz}]$, $\text{F}_4 := [150\,\text{GHz}; 175\,\text{GHz}]$, and $\text{F}_5 := [175\,\text{GHz}; 200\,\text{GHz}]$ are all consecutive sub-intervals in the forbidden frequency range (c.f. Fig. 8).

The results of the performed HYP tissue classifications which are quantified using the introduced metrics are illustrated in Fig. 12. The upper row (a)-(c) shows the results of classification 1 for the identification of the volume fraction $c_\text{v}$ within all frequency intervals $\text{F}_1$ to $\text{F}_5$, whereas the lower row (d)-(f) displays the corresponding results of classification 2 for the cell diameter $d_\text{inc}$. The accuracy, precision and sensitivity of the prediction within classification 1 confirms (as conjectured) that the frequency intervals $\text{F}_1$ and $\text{F}_2$, encompassing predominantly the allowed frequency band below the validity limit are best suited to predict (with measures better than $98\,\%$) aggregate quantities of the microstructure such as the volume fraction $c_v$, as the prediction quality considerably drops (and becomes imbalanced between classified volume fractions) for the residual higher frequency intervals in the forbidden frequency range. This conclusion is also supported by the analysis of classification 2, which aims to identify the cell diameter $d_{inc}$. Given the particular parameter sets the frequency intervals $F_2$ - $F_4$ or $F_3$ and $F_4$ (within the forbidden frequency range) turned out to be well adapted for the prediction of the microscopic quantities of the underlying microstructure. From the considerably strong predictive power associated to the frequency interval $F_2$, we conjecture that the very transition region around the validity limit encompassing both allowed and forbidden frequency ranges could be particularly fertile with respect to potential information on the underlying microstructure. For this reason, the next section addresses how to identify frequency domains that are relevant for the classification of specific microscopic properties.

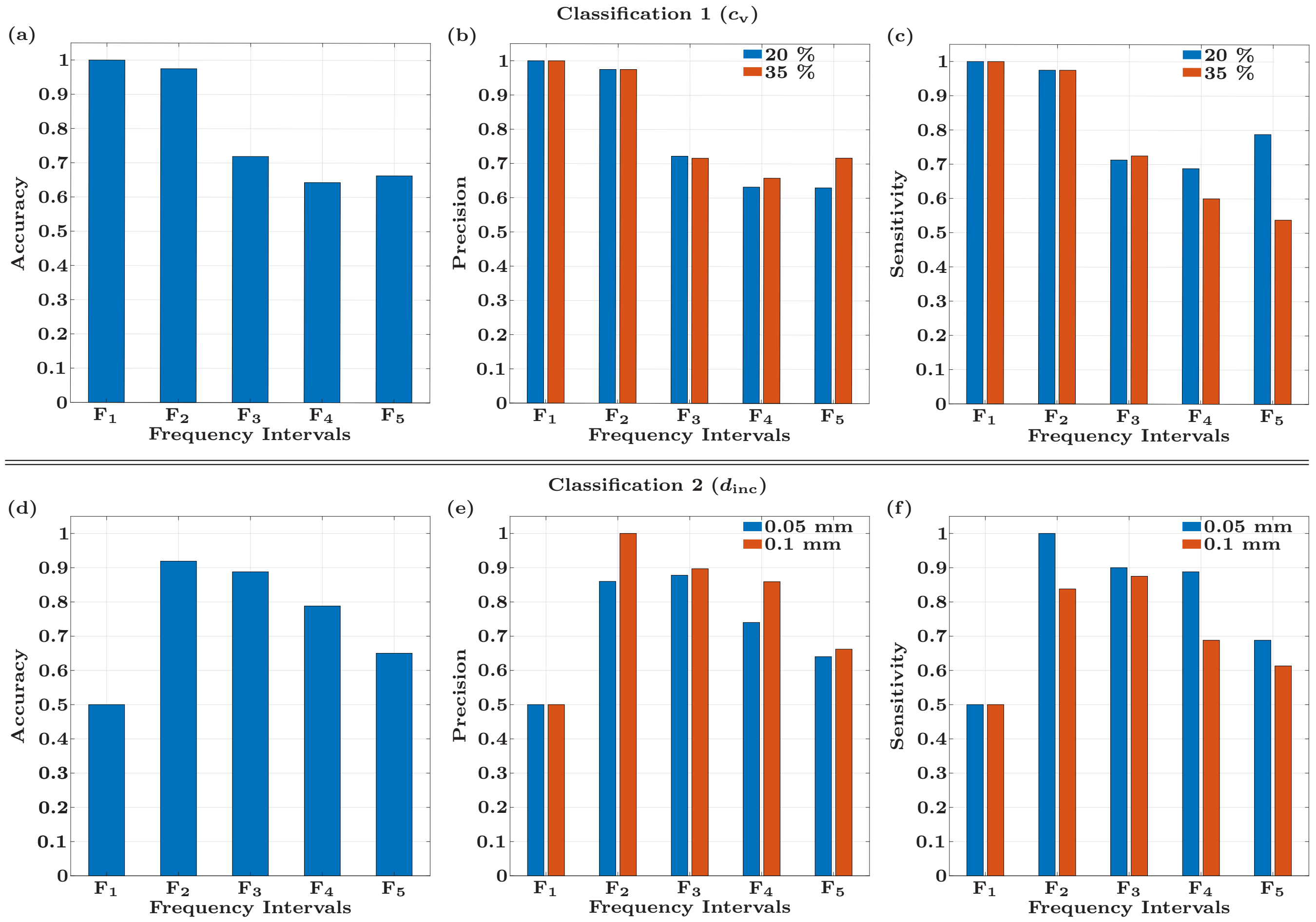

Fig. 12. Results of the material classification based on the implemented binary classification scheme:: The top row, (a) to (c), shows the classification results of the neural networks which had the aim of identifying the volume fraction and the bottom row, (d) to (f), shows the classification results with the aim of identifying the inclusion diameter. The diagrams on the left-hand side show the accuracy of the characterizations, these in the middle the precision and those on the right-hand side the sensitivity of the classifications within the frequency ranges $F_1$ to $F_5$.

## VI. Optimization of the Classification Based on Occlusion Sensitivity

The emerging question is how to separate frequency intervals which bear specific information about the tissue's microstructure from those which do not. To answer this question, we have developed a method based on occlusion sensitivity [25], which we will call occlusion sensitivity method (OSM) in the following. The basic idea is to cover parts of an image with a mask and quantify its influence on object recognition. The method developed will be explained using the data for the classification of the inclusion diameter in interval $F_2$ as an example.

A detailed overview of how the OSM is implemented and modified to separate frequency intervals which are important from those which are not is illustrated in Fig. 13(a). In a first step, the spectral fingerprint of a single microstructure (still represented by an 1D grayscale image) is reproduced in its entirety as many times as the number of pixels it contains, in this case 300. Then, these copies are grouped into a batch and modified by a mask of arbitrary odd pixel length placed in a position shifted by one pixel in each subsequent copy. To enable the mask to hide the information content of the covered pixels without having an over-weighted influence on the pattern recognition of the ANN, the gray value of the mask corresponds to the average value of the grayscale image under consideration.

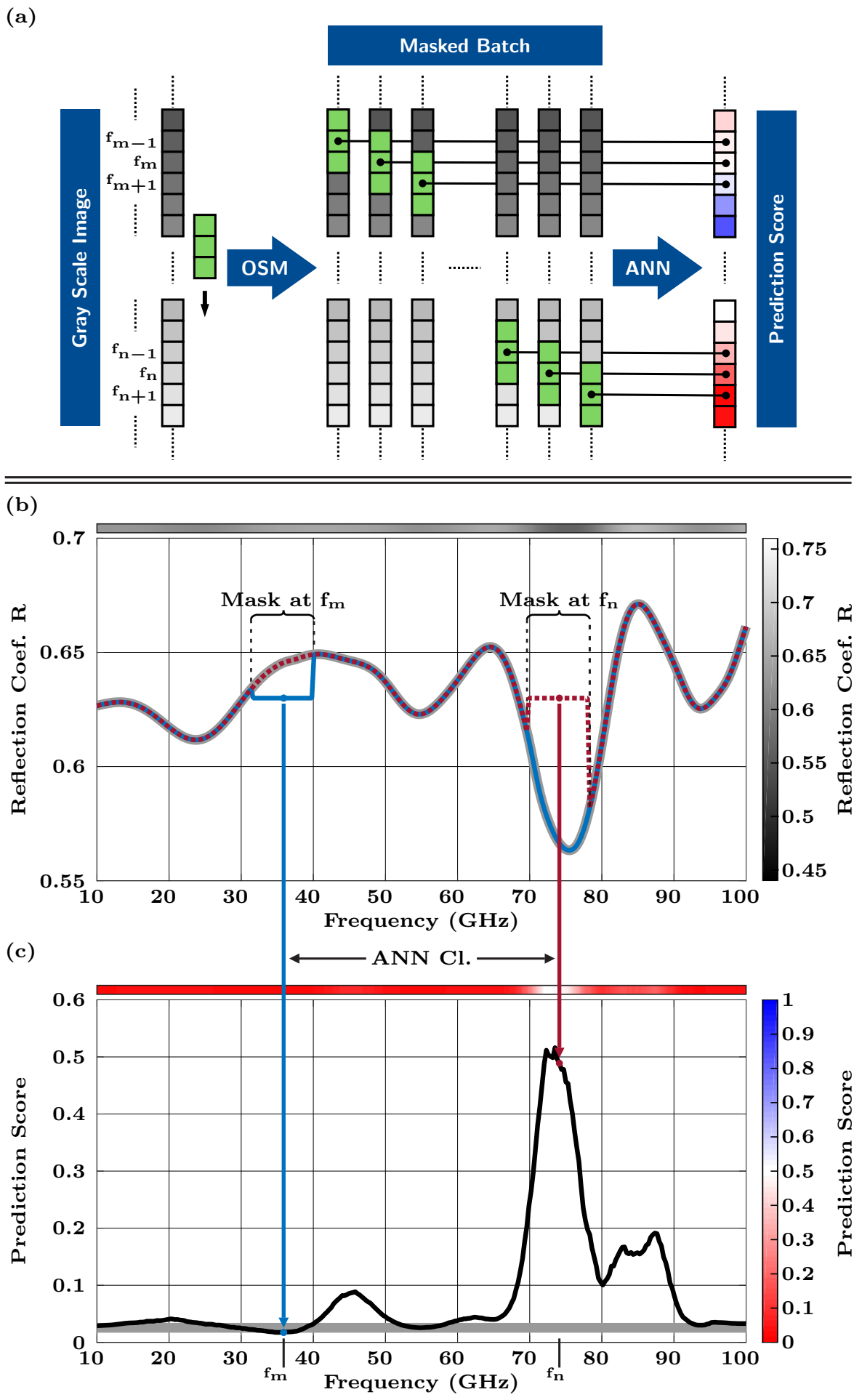

Fig. 13. Bsic idea of the developed OSM: (a) Schematic of the proposed procedure for the separation of information-carrying frequency ranges for the optimization of the frequency specific classification of specific structural/morphological tissue properties. (b) The reflection coefficient of one realization with a volume coefficient $c_v = 0.2$ and inclusion diameter $d_{inc} = 0.1\,\text{mm}$ between 10 to 100 GHz. The gray bar above displays the reflection coefficient of this particular realization as a 1D grayscale image. The color bar associated to this code is located to the right of the graph. The blue dotted and red dashed line show reproductions of the original reflection coefficient modified by masks. The center of the mask is centrally aligned with the sampled frequency points $f_n$ and $f_m$, respectively. (c) depicts the precision score of the modified reproductions (such as the blue dotted and red dashed curve). The prediction score at each frequency point is associated to the center position of the corresponding mask. The gray constant represents the prediction score of the realizations original reflection coefficient. The red bar above the graph displays the prediction score in accordance to the color bar to the right of the diagram.

Finally, the masked batch is transferred to the ANN trained in Sec. V-B to predict the inclusion diameter between 10 and 100 GHz ($F_2$) and the influence of this masking process to the classification is determined. This is done by observing the change in the classification result calculated by the output layer of the ANN, i.e. the softmax function. In the case of our binary classification, the output of the softmax function, the precision score, is a number between 0 and 1, which expresses the probability of belonging to a certain class. A prediction score >0.5 (shaded in blue) means that the ANN has interpreted the spectral response as originating from a microstructure with an inclusion diameter of $d_{inc,1} = 0.05\,\text{mm}$ and a prediction score < 0.5 (shaded in red) as originating from a material structure with an inclusion diameter of $d_{inc,2} = 0.1\,\text{mm}$. An example of how this procedure is actually implemented is given n Fig. 13(b) and (c), using the spectral response of a realization with a volume coefficient of $c_v = 0.2$ and an inclusion diameter of $d_{inc} = 0.1\,\text{mm}$. The coded 1D grayscale image of this particular realization is located above the corresponding curve of the reflection coefficient and represents the input data within the schematic shown in Fig. 13(a). The dotted blue and dashed red curves correspond to reproductions of the original spectral response, where the masks are centered at the frequency points $f_m$ and $f_n$ respectively (cf. masked batch in Fig. 13(a)). The curve of the prediction score in Fig. 13(c), can be determined by the sequential classification of all masked reproductions by the ANN (cf. prediction score in Fig. 13(a)). In analogy to the grayscale coding of the spectral fingerprint, the prediction score is also color-coded and displayed as a red 1D image placed above the graph. The prediction score at each frequency point is associated to the center position of the corresponding mask, illustrated by the arrows connecting Fig. 13(b) and (c). To assess the influence of the mask relative to the prediction score of the original (unmasked) fingerprint, it is also plotted as a gray-shaded constant.

If this procedure is not applied to a single implementation but to the spectral response of all 160 classified realizations of the test set between 10 and 100 GHz, a 2D grayscale image is obtained as shown in Fig. 14(a). In this grayscale image, every pixel in the horizontal direction represents one realization and every pixel in the vertical direction the power reflection coefficient, $R$, at a certain frequency (10 to 100 GHz). The shade varies from black for the lowest $R$ in this frequency range (approximately 0.45) to white for the highest $R$ (approximately 0.75). To support the reference to the procedure presented in Fig. 13, this grayscale image is illustrated in the adjacent figure in simplified form as a pixel map (1). After classification, the spectral response of each realization is ordered from left to right according to its prediction score, displayed as a narrow bar below the grayscale image (2). The color bar associated with the prediction score is placed next to Fig. 14(b). Realizations with an inclusion diameter of $d_{inc,1} = 0.05\,\text{mm}$ which have been correctly identified are labeled as $d^{true}_{inc,1}$ and those with an inclusion diameter of $d_{inc,2} = 0.1\,\text{mm}$ are labeled as $d^{true}_{inc,2}$ or $d^{false}_{inc,1}$, respectively, if the classification is correct and incorrect. The adaptation of the batching can be represented for all realizations by a cube (3). The reproductions of the spectral fingerprints batched in this cube are also modified by masks indicated by green voxels. Each of the modified realizations of the batch is then classified by the ANN. The result of this classification is a 2D red-blue image that documents the influence of the masking for each of the considered realizations against the frequency (4). With this procedure applied to the realizations shown in Fig. 14(a) with a mask size of 163 pixels, Fig. 14(b) is obtained. By relating the prediction score of the unmasked spectral finger prints to Fig. 14(b), the deviation between them can be interpreted as the sensitivity (meant literally) of the ANN to the masked interference patterns, so that the importance of these patterns can be assessed. Assessing Fig. 14(b) in this way, we can conclude from the strong misclassification of previously correctly identified realizations with an inclusion diameter of $d_{\text{inc},2} = 0.1\,\text{mm}$ ($d^{\text{true}}_{\text{inc},2}$) that the frequency response between 40 to 100 GHz plays a major role in the correct classification of the material structure by the ANN.

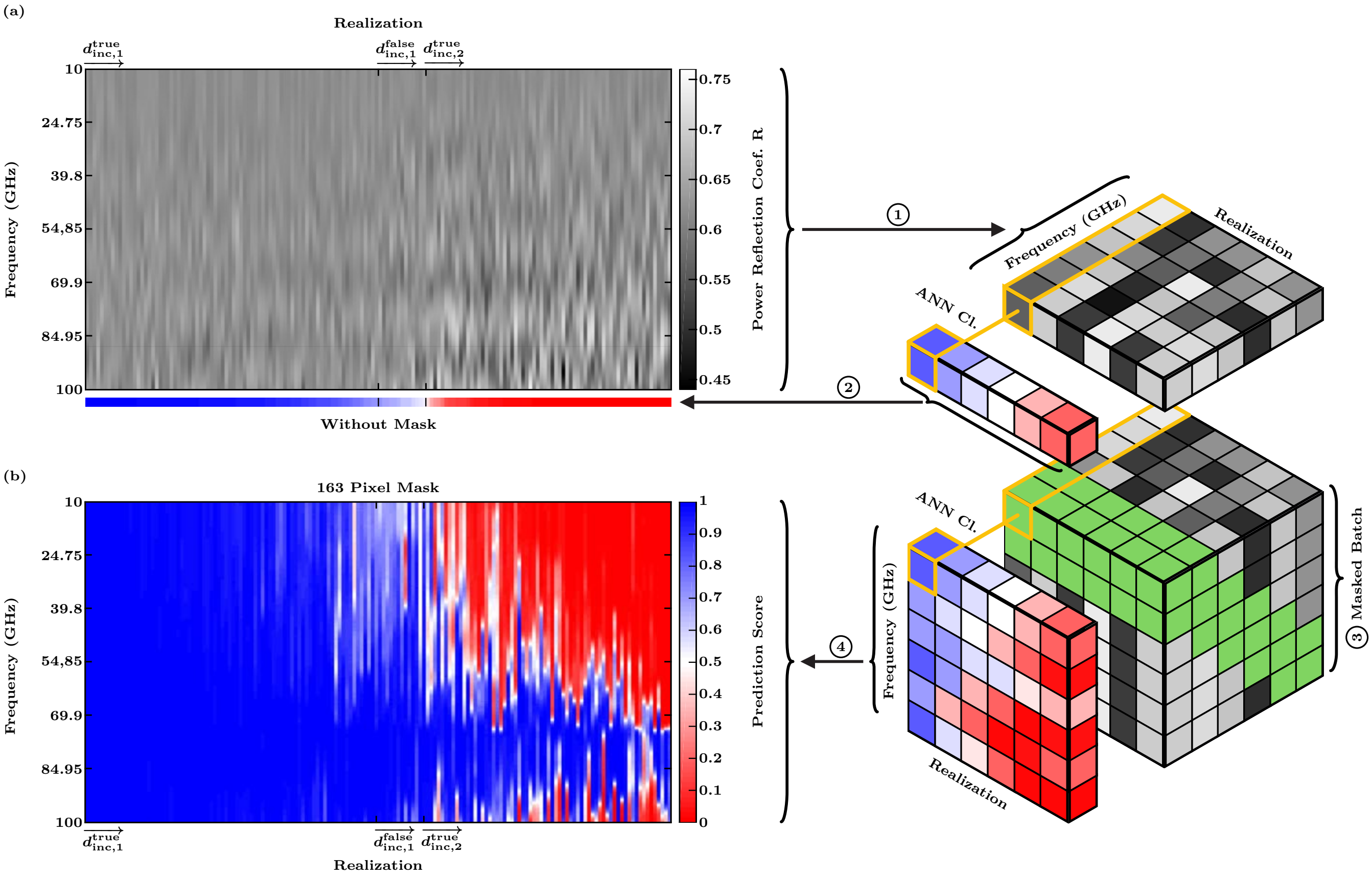

Fig. 14. Application of the OSM to 160 realizations (test set): (a) Grayscale image of 160 realizations. Every pixel in the horizontal direction represents one realization and every pixel in the vertical direction the power reflection coefficient $R$ at a sampled frequency point in the frequency range between 10 to 100 GHz. These spectral responses are ordered from left to right according to its prediction score. $d_{\mathrm{inc},1}^{\mathrm{true}}$ marks realizations correctly assigned to the class $d_{\mathrm{inc},1} = 0.05\,\mathrm{mm}$, $d_{\mathrm{inc},1}^{\mathrm{false}}$ marks those incorrectly assigned to $d_{\mathrm{inc},1} = 0.05\,\mathrm{mm}$ and $d_{\mathrm{inc},2}^{\mathrm{true}}$ marks those correctly assigned to $d_{\mathrm{inc},2} = 0.1\,\mathrm{mm}$. (b) Blue-red image of the prediction score generated by applying the OSM-based optimization procedure to all 160 realizations for a mask size of 163 pixel. The deviation of the prediction score of the original (unmasked) fingerprints with the prediction score of their masked reproductions at a specific frequency point indicates the importance of the frequency patterns covered by this masks.

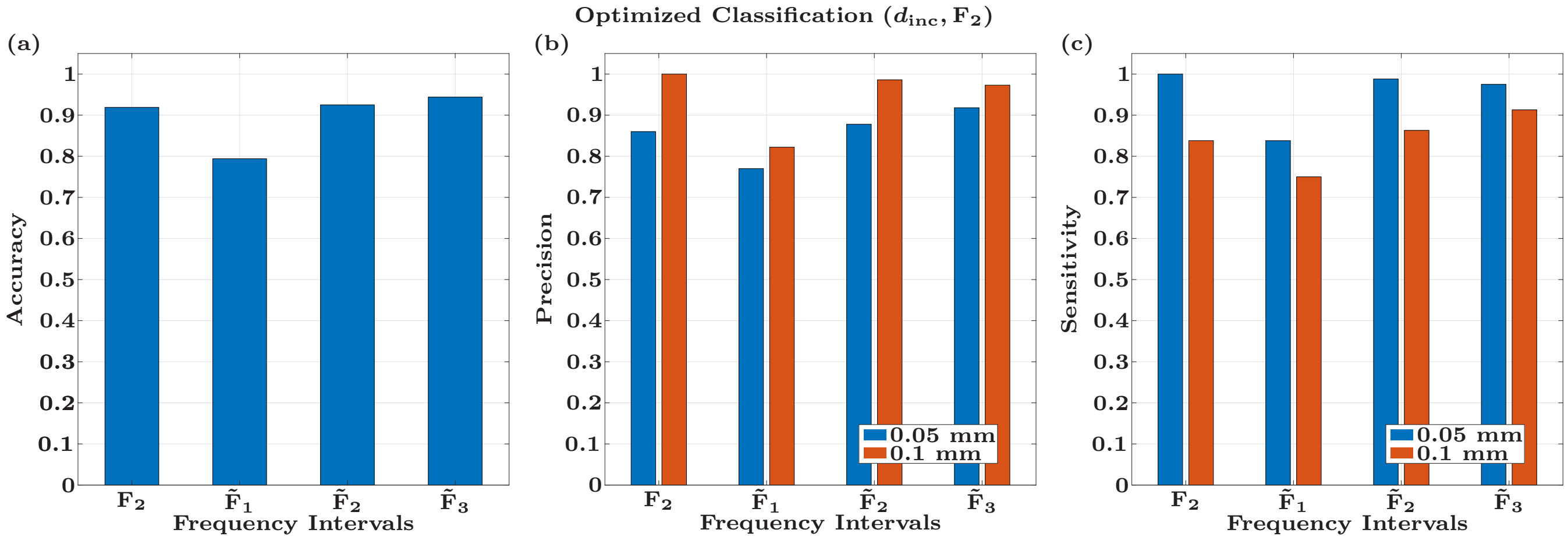

Fig. 15. Cassification results with the aim of identifying the inclusion diameter in the original frequency interval $\mathrm{F}_2$ and in the modified intervals $\tilde{\mathrm{F}}_1$ to $\tilde{\mathrm{F}}_3$. The diagram on the left-hand side shows the accuracy of the characterizations, the one in the middle the precision and the one on the right-hand side, the sensitivity of the classifications.

Therefore, new frequency intervals for optimized classification are defined: $\tilde{\text{F}}_1 := [10\,\text{GHz}; 40\,\text{GHz}]$, $\tilde{\text{F}}_2 := [40\,\text{GHz}; 100\,\text{GHz}]$ and $\tilde{\text{F}}_3 := [100\,\text{GHz}; 110\,\text{GHz}]$.

The results of this optimized classification are depicted in Fig. 15. Comparing the original frequency interval $\text{F}_2$ (from Fig. 12) with the frequency interval $\tilde{\text{F}}_1$, it can be seen that the accuracy decreases from $91.9\,\%$ to $79.4\,\%$. This indicates that the information contained in interval $\tilde{\text{F}}_1$ has decreased. By comparing $\text{F}_2$ with $\tilde{\text{F}}_2$ on the other hand, it can be seen that the prediction accuracy has increased slightly from $91.9\,\%$ to $92.5\,\%$ and in addition the asymmetry in precision and sensitivity has decreased. If $\tilde{\text{F}}_2$ is extended by $10\,\text{GHz}$ to $110\,\text{GHz}$ in $\tilde{\text{F}}_3$, the accuracy can be increased even by $1.9\,\%$ percent and the asymmetry decreases further.

Based on these results it could be shown that the reflection coefficient can be used as a spectral fingerprint containing information about specific morphological/structural properties in specific frequency intervals. Furthermore, based on the OSM, we were able to develop a procedure that selects and visualizes these information-bearing frequency intervals.

Since our study is based on generic, 2D HYP structures only, we will show through 3D simulations that the main statements of this study are also transferable to more realistic scenarios in Sec. VII.

## VII. Summary and Conclusions

In this study, we presented a three-stage methodology in the framework of a parametrizable multiscale EM simulation workbench yielding the effective EM material properties of a (bio-)composite, considering both anisotropy and dispersion, which are rooted in the precise cellular structure of the composite. In a subsequent step these homogenized material models have been evaluated in terms of their validity limit. This limit has been determined as the frequency at which a deviation of $2\,\%$ occurred between the frequency responses of the heterogeneous (bio-)composite (reference model) and its homogeneous representation in a generic reflectometry setup in which both materials are illuminated by an EM plane wave. In our study we focused on a 2D analysis of generic but suitable HYP representations of a complex tissue composite. Since the validity of homogenization at the tissue level is the main focus of the intended multi-scale modeling approach, the HYP representations investigated were modeled as randomized cell arrangements.

The modeling and simulation of a large number of realizations of these HYP structures in Sec. 3 led to a comprehensive Monte Carlo analysis of various generic HYP derivatives. As part of this analysis we developed a statistical measure to determine the validity limit of the implemented quasi-static homogenization approach, which allows us to divide the spectral response of the reflection into a validity range in which the homogenization is valid, and a forbidden range in which it is not. The results of this analysis revealed that the applicability of the homogeneous material representation had an upper operating frequency limit that started at surprisingly low frequencies in the low mm-wave range and thus contradicted the traditional use of homogenized layer models in tissue analysis/diagnostics [12], [16]. This forced any hierarchically organized multiscale model topology to become strongly tied to a corresponding operating frequency bandwidth. Consequently, an ultra-wideband tissue analysis must consider simultaneous structural changes in the tissue models during frequency domain simulation by using different models to represent different frequency ranges.

Sec. IV showed that the collapse of the EMT approximation coincided with an increasing impact of the microscopic structural/morphological properties on the frequency response beyond the validity limit. In the course of this study, the difference in electric and magnetic energy densities displayed particularly strong resonances and standing fields which correlate highly with the tissue microstructure and thus with the randomized HYP cell distribution. To explain the relationship between interference patterns in the spectral response of the reflection coefficient and the underlying microstructure, and thus to establish reflection as a spectral fingerprint to classify structural/morphological tissue properties, we studied specific frequency points in the spectral response of the power reflection R and absorption A in the forbidden range for a single HYP realization in conjunction with associated EM field patterns in the microstructure as an illustrative example. The resulting spectral fingerprint opens up the possibility to identify and classify specific features of the tissue microstructure.

Following this idea, we extended the Monte Carlo analysis described in Sec. 2 by 4 additional parameter sets with variable structural parameters for the inclusion diameter and the volume fraction of the lipid droplets and identified the resulting HYP derivatives (represented by the parameter sets) with a multilabel, image recognition-based material classification procedure using ANN in Sec. 5. Within this classification we were able to correctly assign the tissue derivatives to the correct class on the basis of reflection with high reliability (accuracy up to $89\,\%$) and show that the valid and the forbidden frequency ranges complemented each other with regard to the information content contained in the spectral fingerprint. By re-labeling the underlying data set and modifying the architecture of the ANN to perform a binary classification for the targeted identification of either inclusion diameter or volume fraction, we were able to show that the forbidden frequency range could be exploited to predict the microstructures of the HYP tissues (e.g. the expectation values of the cell sizes) and the validity range to classify aggregated HYP tissue properties such as the volume fractions of the adipose cells. In Sec. 6, we developed the occlusion sensitivity method, not only to improve accuracy, but also to improve the precision and sensitivity in the classification of HYP derivatives by separation and subsequent selection of the information-carrying frequency ranges.

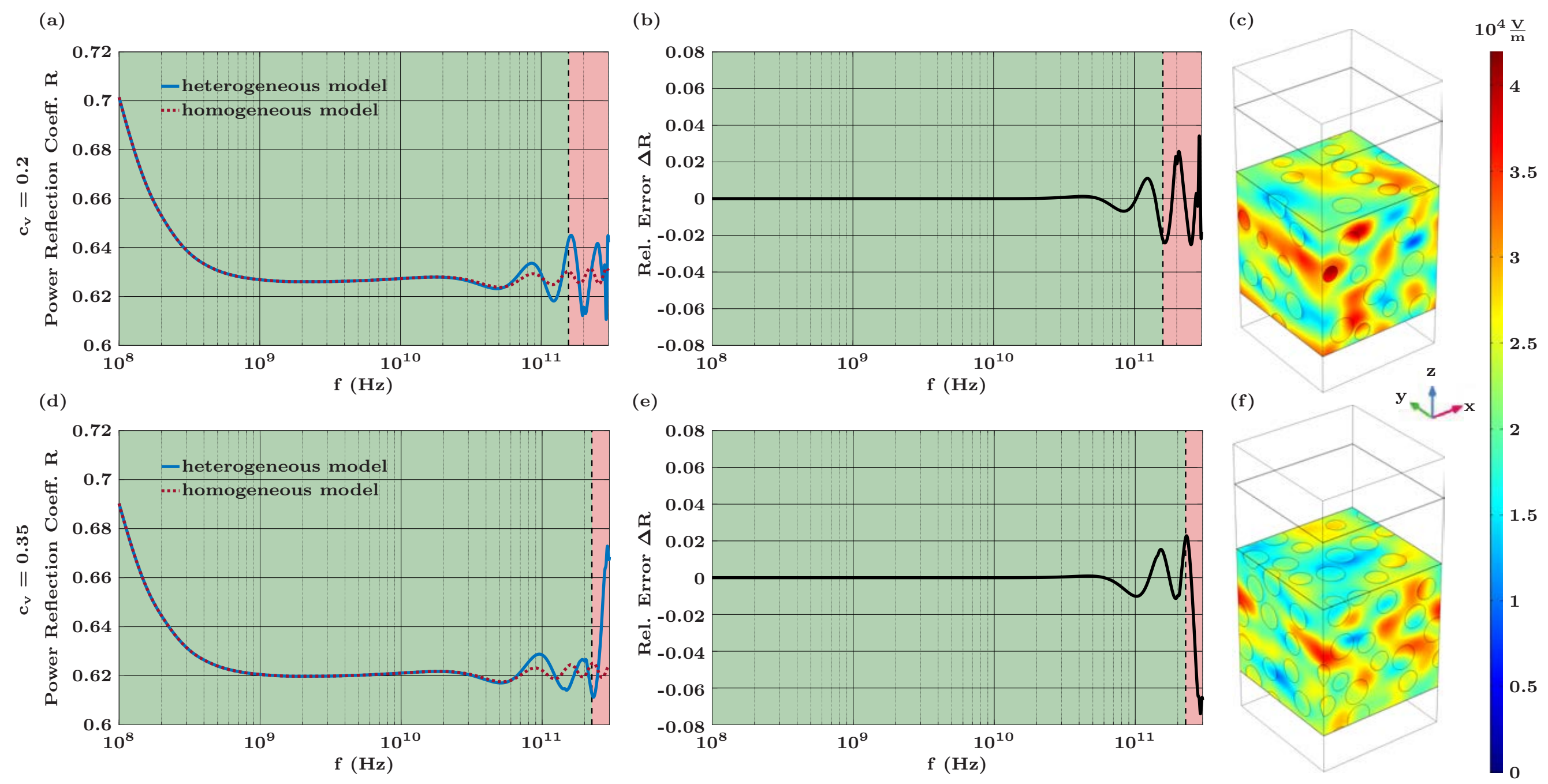

Fig. 16. Overview of two 3D simulations of a generic HYP structure (i.e. for $\varepsilon_{r,1} = 80$; $\sigma_1 = 0.53\,\mathrm{S/m}$; $\varepsilon_{r,2} = 50$; $\sigma_2 = 0.12\,\mathrm{S/m}$; $d_{\mathrm{inc}} = 50\,\mu\mathrm{m}$). The edge length of the simulated supercell is $0.25\,\mathrm{mm}$. The upper row contains simulation results for a structure with a volume fraction of $c_{\mathrm{v}} = 0.2$ and the bottom row those of a structure with a volume fraction of $c_{\mathrm{v}} = 0.35$. (a) and (d) show the frequency-dependent power reflection coefficient $R$ of the heterogeneous model (blue) and the homogeneous (red). (b) and (e) provide an overview of the relative error between both models. The validity range is highlighted in «green» and the forbidden frequency range in «red». (c) and (e) showe the field distribution of the magnitude of the electric field within the supercell embedded in the simulation setup at $300\,\mathrm{GHz}$.

On the basis of these research results, the following priorities emerge for the future: Further metrological research and documentation of validity limits caused by structural/morphological signatures in the frequency response of reflection in the forbidden frequency range will be carried out. For this purpose, specially designed 2D composites will be created using 3D printing and used to make measurements for comparison with corresponding simulations to explore the EM signatures in the forbidden range for composites with low material contrast. In addition, structures with embedded channels filled with liquids (e.g. water, alcohol) will be examined to investigate the separation of signatures originating in the morphology/structure and those in the dispersive EM properties of the constituents of the composite. A further future focus will be the expansion of the virtual workbench for the exploration of multi-layer structures. The central question is how much the structural/morphological signatures of the individual layers can be superimposed in the frequency response and still be separated in terms of the allowed operating frequency bands of the EMT models of the individual layers as expressed by their validity limits. The aim will be to identify and analyze sub-frequency bands in order to extract information from individual layers in isolation and to deduce a complete image of the material system using ANN. Furthermore, the question arises to what extent the roughness of the material surface has an influence on the analysis of a composite. This question can be divided into two parts: firstly, whether volume and surface scattering can be separated and secondly, whether the type of discretization has an influence on the result when simulating reflection events at a rough material boundary layer.

## Appendix: 3D Reference Simulation

The transferability of the research results is checked by simulations on two three-dimensional microstructures, one with a volume coefficient of $c_{\mathrm{v}} = 0.2$ and the other with $c_{\mathrm{v}} = 0.35$. As shown in Fig. 16(c) and (f), microstructures with spherical inclusions embedded in a cubic supercell are investigated. While the material parameters and the diameter of the inclusion, $d_{\mathrm{inc}} = 0.05\,\mathrm{mm}$, remain unchanged to those introduced in Sec. II, the edge length of the cubic supercell is now reduced to $0.25\,\mathrm{mm}$. To numerically relieve the simulation, an auxiliary layer has been inserted between the supercell and the PML. Just like the bottom PML, the effective material parameters of the composite are assigned to this auxiliary layer. Note that this separation of the heterogeneous layer's supercell in the 3D simulations by this auxiliary layer is mainly due to numerical reasons as the direct termination of the heterogeneous layer by the PML poses some numerical challenges given the heterogeneous layer's complex morphology.

Due to the high numerical demand and, thus, to the high simulation time, this investigation could not be performed with a larger set of implementations. The simulations has been performed on PC equipped with two Intel Xeon E5-2697 V4 processors (36 cores) and 512 GB of RAM. Between 0.1 and 300 GHz, 200 frequency points have been simulated with a MUMPS solver whereby the density of the frequency points for higher frequencies has increased. The simulation for each frequency point lasts 2 hours and includes the full wave simulation of the heterogeneous microstructure and its homogenized representative in p- and s-polarization.

The spectral response of the power reflection coefficient $R$ is shown for the heterogeneous (blue) and homogeneous model (red curve) in the given frequency interval in Fig. 16(a) and (d). Based on the $2\,\%$ deviation criterion, the validity of the homogeneous material model is retrieved from their relative deviation (i.e. relative error) and collapses at 156 GHz for the material structure with a volume coefficient of $c_\mathrm{v} = 0.2$ and at 225 GHz for $c_\mathrm{v} = 0.35$. The validity range of the homogeneous model is again marked green and the forbidden range red.

The observed periodic oscillations in the simulated spectral response of the power reflection are due to Fabry-Pérot resonances in both, the heterogeneous and homogeneous layer. This oscillations are determined by finite thickness of the layers and thus not present in our prior 2D simulations as we have dealt there with a numerically infinite half-space system. However, the evaluation of the validity limit is not affected by the Fabry-Perot resonances since they are present in both models and thus assumed to be balanced out in the analysis of the relative error. To underline this, Fig. 16(b) and (e) show the relative errors between the reflection coefficients of both models in the same frequency range.

The field strength distributions within the heterogeneous models at 300 GHz illustrated in Fig. 16(c) and (f) show that the breakdown of the validity of the EMT model and the associated interference patterns in the reflection response are accompanied by strong field inhomogeneities within the microstructure. Thus the transferability of the research results to three-dimensional microstructures is clearly indicated.

## Acknowledgment

The authors also thank Mr. Viktor Gerhardt and Mr. Knut Mäß for their excellent technical support, the student assistants Mrs. Katja Ehlen, Mr. Joey Maibaum and Mr. Jannis Assmann for their reliable and fast work and Mr. Barry Morley for his thorough proofreading.

## References

[1] P. Hillger, M. van Delden, U. S. M. Thanthrige, A. M. I. Ahmed, J. Wittemeier, K. Arzi, B. S. M. Andre, W. Prost, A. Rennings, D. Erni, T. Musch, N. Weimann, A. Sezgin, N. Pohl, and U. R. Pfeiffer, "Toward mobile integrated electronic systems at THz frequencies," *J. Infrared Millim. Terahertz Waves*, 2020.

[2] J. Barowski, M. Zimmermanns, and I. Rolfes, "Millimeter-wave characterization of dielectric materials using calibrated FMCW transceivers," *IEEE Trans. Microwave Theory Techn.*, vol. 66, no. 8, pp. 3683–3689, 2018.

[3] J. Barowski, J. Jebramcik, I. Alawneh, F. Sheikh, T. Kaiser, and I. Rolfes, "A compact measurement setup for in-situ material characterization in the lower THz range," in *2nd Int. Workshop on Mobile THz Syst. (IWMTS 2019)*, Bad Neuenahr, Germany, 2019.

[4] J. Dong, B. Kim, A. Locquet, P. McKeon, N. Declercq, and D. S. Citrin, "Nondestructive evaluation of forced delamination in glass fiber-reinforced composites by terahertz and ultrasonic waves," *Compos. B. Eng.*, vol. 79, pp. 667–675, 2015.

[5] J. Dong, A. Locquet, N. F. Declercq, and D. S. Citrin, "Polarization-resolved terahertz imaging of intra- and inter-laminar damages in hybrid fiber-reinforced composite laminate subject to low-velocity impact," *Compos. B. Eng.*, vol. 92, pp. 167–174, 2016.

[6] L. Afsah-Hejri, P. Hajeb, P. Ara, and R. J. Ehsani, "A comprehensive review on food applications of terahertz spectroscopy and imaging," *Compr. Rev. Food Sci. Food Saf.*, vol. 18, no. 5, pp. 1563–1621, 2019.

[7] X. Hu, W. Lang, W. Liu, X. Xu, J. Yang, and L. Zheng, "A non-destructive terahertz spectroscopy-based method for transgenic rice seed discrimination via sparse representation," *J. Infrared Millim. Terahertz Waves*, vol. 38, no. 8, pp. 980–991, 2017.

[8] F. Toepfer and J. Oberhammer, "Millimeter-wave tissue diagnosis: The most promising fields for medical applications," *IEEE Microw. Mag.*, vol. 16, no. 4, pp. 97–113, 2015.

[9] D. Oppelt, J. Adametz, J. Groh, O. Goertz, and M. Vossiek, "MIMO-SAR based millimeter-wave imaging for contactless assessment of burned skin," in *2017 IEEE MTT-S Int. Microw. Symp. (IMS 2017)*, Honolulu, HI, USA, 2017.

[10] X. Yang, X. Zhao, K. Yang, Y. Liu, Y. Liu, W. Fu, and Y. Luo, "Biomedical applications of terahertz spectroscopy and imaging," *Trends Biotechnol.*, vol. 34, no. 10, pp. 810–824, 2016.

[11] K. I. Zaytsev, I. N. Dolganova, N. V. Chernomyrdin, G. M. Katyba, A. A. Gavdush, O. P. Cherkasova, G. Komandin, M. A. Shchedrina, A. N. Khodan, D. S. Ponomarev, I. V. Reshetov, V. Karasik, M. Skorobogatiy, V. N. Kurlov, and V. V. Tuchin, "The progress and perspectives of terahertz technology for diagnosis of neoplasms: A review," *J. Opt.*, vol. 22, no. 1, pp. 1–44, 2019.

[12] S. I. Alekseev and M. C. Ziskin, "Human skin permittivity determined by millimeter wave reflection measurements," *Bioelectromagnetics*, vol. 28, no. 5, pp. 331–339, 2007.

[13] S. Huclova, D. Erni, and J. Froehlich, "Modelling effective dielectric properties of materials containing diverse types of biological cells," *J. Phys. D: Appl. Phys.*, vol. 43, no. 36, pp. 365 405–1–10, September 2010.

[14] ——, "Modelling and validation of dielectric properties of human skin in the MHz region focusing on skin layer morphology and material composition," *J. Phys. D: Appl. Phys.*, vol. 45, no. 2, pp. 025 301–1–17, January 2012.

[15] J. Froelich, S. Huclova, and D. Erni, *book chapter 12 "Accurate multi-scale skin model suitable for determining sensitivity and specificity of changes of skin components," pp. 353-394, in Computational Biophysics of the Skin, Bernard Querleux, Ed.*, Singapore: Pan Stanford Publishing Pte. Ltd., 2014.

[16] M. Saviz, L. Mogouon Toko, O. Spathmann, J. Streckert, V. Hansen, M. Clemens, and R. Faraji-Dana, "A new open-source toolbox for estimating the electrical properties of biological tissues in the terahertz frequency band," *J. Infrared Millim. Terahertz Waves*, vol. 34, no. 9, pp. 529–538, 2013.

[17] O. Spathmann, M. Zang, J. Streckert, V. Hansen, M. Saviz, T. M. Fiedler, K. Statnikov, U. R. Pfeiffer, and M. Clemens, "Numerical computation of temperature elevation in human skin due to electromagnetic exposure in the thz frequency range," *IEEE Trans. THz Sci. Technol.*, vol. 5, no. 6, pp. 978–989, 2015.

[18] K. Jerbic, B. Sievert, J. T. Svejda, A. Rennings, and D. Erni, "On the applicability of homogenization in composite material models for tisse analysis in the mm-wave range," in *Photon. & Electromagn. Res.*

*Symp./Prog. in Electromagn. Res. (PIERS 2019)*, University of Rome 'La Sapienza', Rome, Italy, 2019.

[19] S. Huclova, "Modeling of cell suspensions and biological tissue for computational electromagnetics," Ph.D. dissertation, ETH Zurich, Zurich, Switzerland, 2011.

[20] H. Richter, "Mote3d: an open-source toolbox for modelling periodic random particulate microstructures," *Model. Simul. Mat. Sci. Eng.*, vol. 25, no. 3, p. 035011, 2017.

[21] I. Krakovský and V. Myroshnychenko, "Modeling dielectric properties of composites by finite-element method," *J. Appl. Phys.*, vol. 92, no. 11, pp. 6743–6748, 2002.

[22] *COMSOL Multiphysics®* (v. 5.4), COMSOL AB, Available: www.comsol.com.

[23] G. Solomakha, J. T. Svejda, C. van Leeuwen, A. Rennings, A. J. Raaijmakers, S. Glybovski, and D. Erni, "A self-matched leaky-wave antenna for ultrahigh-field MRI with low SAR," *arXiv:2001.10410 [physics.app-ph], Bibcode: 2020arXiv200110410S*, 2020.

[24] B. Cabral and L. C. Leedom, "Imaging vector fields using line integral convolution," in *Proc. of the 20th Ann. Conf. on Comput. Graph. and Interactive Techn.*, M. C. Whitton, Ed. New York, NY: ACM, 1993, pp. 263–270.

[25] S. Eddins, "Network visualization based on occlusion sensitivity: Mathworks blogs," (Accessed: 28.01.2019). [Online]. Available: https://blogs.mathworks.com/deep-learning/2017/12/15/network-visualization-based-on-occlusion-sensitivity/

## Biographies

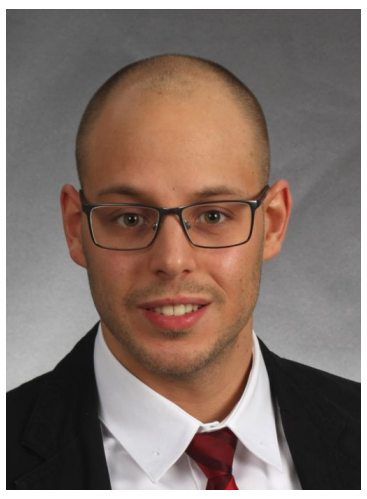

**Kevin Jerbic** received his B.Sc. and M.Sc. in Electrical Engineering from the University of Duisburg-Essen in 2015 and 2017, respectively. Since 2017 he has been working as a member of the department of General and Theoretical Electrical Engineering at the University of Duisburg-Essen in Duisburg, Germany. His current research interests is the development of model based calibration schemes for the classification of material properties of complex material systems using machine learning approaches based on artificial neural networks at mm-wave frequencies. This includes the exploration of validity limits of hierarchical multiscale approaches for numerical homogenization to capture the dispersive and tensorial macroscopic electromagnetic properties of (bio-)composites and the quantification of the contribution of rough interfaces to subsurface scattering.

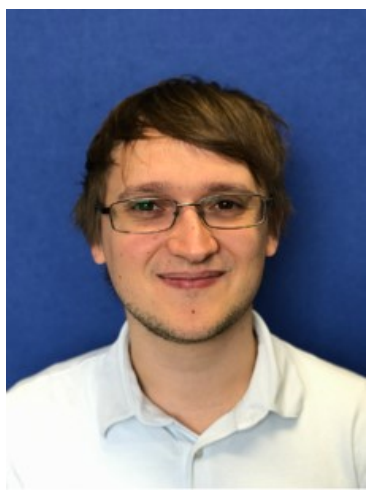

**Kevin Neumann** received his B.Sc. in Nanoengineering in 2015 and his M.Sc. in Power Engineering in 2017, both at the University of Duisburg-Essen. Since 2017, he is working at the DFG project "Flexible Radio Frequency Identification Tags and System (FlexID)" as a Ph.D. student at the department of General and Theoretical Electrical Engineering (ATE) in Duisburg, Germany. His main research focus lies on the full system development of chipless RFID tag, which contains semiconductor modeling as well as electromagnetic wave simulations. Furthermore, he investigates the performance of numerical crumpled antenna structures using a combination of finite element and boundary element method modeling.

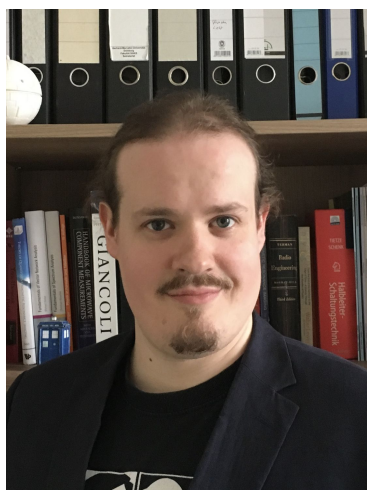

**Jan Taro Svejda** started his electrical engineering career at the University of Applied Science, Düsseldorf, Germany, where he received his B.Sc. degree in 2008. Consecutively he continued his studies in Electrical Engineering and Information Technology at the University of Duisburg-Essen, Duisburg, Germany, and received his M.Sc. degree in 2013 and his Dr.-Ing. Degree in 2019 for his research work in the field of X-nuclei based magnetic resonance imaging, respectively. He is currently working as a research assistant at University of Duisburg-Essen in the department of General and Theoretical Electrical Engineering where he is involved in teaching several lectures and courses mainly in the field of electrical engineering. His general research interest includes all aspects of theoretical and applied electromagnetics, currently focusing on medical applications, electromagnetic metamaterials, and scientific computing methods.

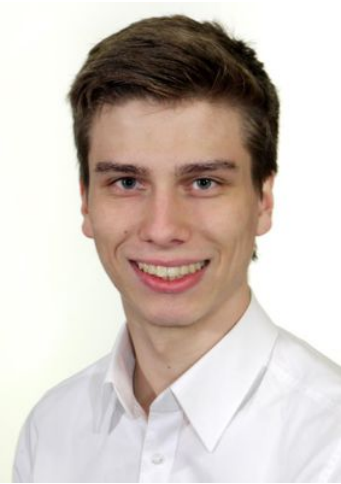

**Benedikt Sievert** was born in Krefeld, Germany. He received his B.Sc. and M.Sc. in Electrical Engineering/High Frequency Systems from the University of Duisburg-Essen in 2017 and 2019, respectively. Since 2017 he is a member of the Laboratory of General and Theoretical Electrical Engineering of the University of Duisburg-Essen. His current research interests include mmwave on-chip antennas, electromagnetic metamaterials, theoretical and computational electromagnetics.

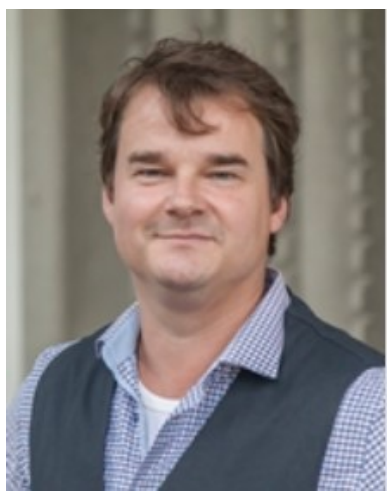

**Andreas Rennings** studied electrical engineering at the University of Duisburg-Essen, Germany. He carried out his diploma work during a stay at the University of California in Los Angeles. He received his Dipl.-Ing. and Dr.-Ing. degrees from the University of Duisburg-Essen in 2000 and 2008, respectively. From 2006 to 2008 he was with IMST GmbH in Kamp-Lintfort, Germany, where he worked as an RF engineer. Since then, he is a senior scientist and principal investigator at the Laboratory for General and Theoretical Electrical Engineering of the University of Duisburg-Essen. His general research interests include all aspects of theoretical and applied electromagnetics, currently with a focus on medical applications and on-chip millimeter-wave/THz antennas. He received several awards, including a student paper price at the 2005 IEEE Antennas and Propagation Society International Symposium and the VDE-Promotionspreis 2009 for the dissertation.

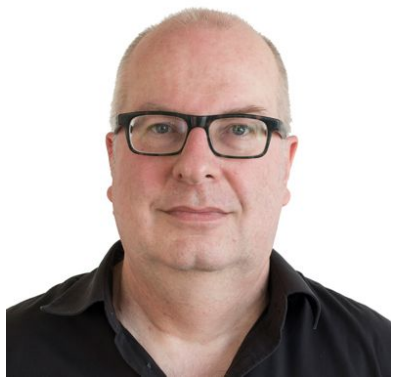

**Daniel Erni** received a diploma degree from the University of Applied Sciences in Rapperswil (HSR) in 1986, and a diploma degree from ETH Zürich in 1990, both in electrical engineering. Since 1990 he has been working at the Laboratory for Electromagnetic Fields and Microwave Electronics, ETH Zürich, where he got his Ph.D. degree in 1996. From 1995-2006 he has been the founder and head of the Communication Photonics Group at ETH Zürich. Since Oct. 2006 he is a full professor for General and Theoretical Electrical Engineering at the University of Duisburg-Essen, Germany (http://www.ate.uni-due.de/). His current research includes nanophotonics, plasmonics, optical antennas, as well as advanced solar cell concepts, optical and electromagnetic metamaterials (e.g. for multi-functional leaky wave antennas and for advanced RF systems for high-field MRI imaging), microwave engineering and THz modeling (e.g. for chipless RFID tags and for THz material and surface characterization), computational electromagnetics, and bioelectromagnetics (e.g. biological tissue modeling). On the system level Daniel Erni has pioneered the introduction of numerical structural optimization into dense integrated optics device design. Further research interests include science and technology studies (STS) as well as the history and philosophy of science with a distinct focus on the epistemology in engineering sciences. He is a Fellow of the Electromagnetics Academy, a member of the Center for Nanointegration Duisburg-Essen (CENIDE), as well of Materials Chain, the Flagship Program of the niversity Alliance Ruhr, and member of the Swiss Physical Society (SPS), the German Physical Society (DPG), the Optical Society of America (OSA), Electrosuisse, and of the IEEE.